\documentclass[twocolumn]{aastex631}
\usepackage{xurl}
\usepackage[section]{placeins}
\usepackage{rotating}
\usepackage{pgfplotstable}
\usepackage[T1]{fontenc}
\usepackage{tgbonum}
\usepackage{appendix}
\usepackage{natbib}

\newcommand{\msun}{M_\odot}

\begin{document}
\title{Model Choice Matters for Age Inference on the Red Giant Branch}

\author[0000-0002-1333-8866]{Leslie M. Morales}
\affiliation{University of Florida, Department of Astronomy, Gainesville, FL, 32611 USA}
\affiliation{San Diego State University, Department of Astronomy, San Diego, CA, 92182 USA}
\author[0000-0002-4818-7885]{Jamie Tayar}
\affiliation{University of Florida, Department of Astronomy, Gainesville, FL, 32611 USA}
\author[0000-0002-9879-3904]{Zachary R. Claytor}
\affiliation{University of Florida, Department of Astronomy, Gainesville, FL, 32611 USA}
\affiliation{Space Telescope Science Institute, 3700 San Martin Drive, Baltimore, MD 21218, USA}

\begin{abstract}
Galactic archaeology relies on accurate stellar parameters to reconstruct the galaxy's history, including information on stellar ages. While the precision of data has improved significantly in recent years, stellar models used for age inference have not improved at a similar rate. In fact, different models yield notably different age predictions for the same observational data. In this paper, we assess the difference in age predictions of various widely used model grids for stars along the red giant branch. Using open source software, we conduct a comparison of four different evolution grids and we find that age estimations become less reliable if stellar mass is not known, with differences occasionally exceeding $80\%$. Additionally, we note significant disagreements in the models' age estimations at non-solar metallicity. Finally, we present a method for including theoretical uncertainties from stellar evolutionary tracks in age inferences of red giants, aimed at improving the accuracy of age estimation techniques used in the galactic archaeology community.
\end{abstract}

\section{Introduction} \label{sec:intro}
Stellar ages provide valuable insight into a multitude of studies beyond stellar evolution, such as galactic evolution and planet formation. The ages of galactic globular clusters (GCs), for instance, have been used to constrain the formation and early evolution of the Milky Way. The very oldest clusters offer clues to the earliest phases of star formation, while the ages of old, but non-primordial GCs have been linked to past accretion events (\citealt{2009Sarajedini}; \citealt{2013VandenBerg}; \citealt{2021Lucertini}; \citealt{2024Valenzuela}).
Galactic astronomers use stellar ages to infer physical properties of galaxies as they evolve (\citealt{2016BlandHawthorn}; \citealt{2017Thomas}; \citealt{2018NatAs...2..483V}). More recently, exoplanet research has begun correlating the ages of host stars with Hot Jupiter formation (\citealt{2013Schlaufman}; \citealt{2022Hamer}; \citealt{2023PNAS..12004179C}). In particular, Galactic Archaeology,  the study of our galaxy’s structure and history, heavily relies on stellar ages to provide accurate timelines on the evolution of our Universe (\citealt{2010Soderblom}; \citealt{2021PhDT........45S}). 

Ages cannot be measured directly. Instead, we infer them from theory or by comparing to established results using standard age indicators. Various methods for age inference are continuously being developed, including: inferring age from [C/N] abundances (\citealt{2015Salaris}; \citealt{2017Lagarde}; \citealt{2019Casali}), using the relation between stellar rotation and time due to magnetic braking (known as gyrochronology) (\citealt{1972Skumanich}; \citealt{2003Barnes}; \citealt{2007Barnes}; \citealt{2020Casal}), and determining kinematic ages based on the motion of stars through space (\citealt{2004Nordstrom}; \citealt{2009Aumer}; \citealt{2018YuLiu}; \citealt{2019Ting}; \citealt{2021Lu}). While diverse in approach, these methods are still inherently reliant on calibrations using stellar models, and therefore remain model-dependent. Most techniques make use of common theoretical frameworks, particularly isochrones and evolutionary tracks. Isochrones model the evolution of stars as a population for a singular age, while evolutionary tracks model the evolution of a single star of a particular mass. Isochrone fitting, a method of estimating stellar ages by comparing observable properties of stars to theoretical models, has been shown to work well when using a star cluster's Hertzsprung-Russel (HR) or color-magnitude diagram (\citealt{2010Dotter}; \citealt{2014Chen}; \citealt{2023Gontrcharov}; \citealt{2024Reyes}). However, this method presents challenges for individual stars on the red giant branch, where closely spaced isochrones make precise age determinations difficult if metallicity is not precisely known. Stars with different masses and metallicities can populate the same narrow region of the red giant branch (\citealt{2002Salaris}; \citealt{2005Gallart}; \citealt{2018Valle}; \citealt{2021SahlholdtLindegren}).

In practice, age estimation methods often make use of stars along the subgiant and red giant branch (SGB, RGB) because they are visible over large distances and in great numbers. As stars move off the main sequence to the subgiant phase, age can be directly inferred from the luminosity (\citealt{2021ApJ...915...19G}; \citealt{2022XiangRix}). This is because luminosity of the SGB is strongly dependent on the mass of the stellar core, which scales with the total mass and as a result, tightly correlated with age (\citealt{2005Salarisbook}; \citealt{2013Cassisibook}; \citealt{2013Kippenhahn}; \citealt{2024Nataf}). While these stars are useful, they are visible only in limited regions of the galaxy and at shorter distances. In contrast, red giants, being significantly brighter, can be observed at greater distances. For red giants, we can infer an age from the metallicity and mass of the star. However, along with the importance of accurate metallicity measurements mentioned previously, mass is not always available and the ability to infer mass from the effective temperature is difficult, although occasionally possible with a combinations of spectroscopic, photometric, or asteroseismic data (\citealt{2016Feuillet}; \citealt{2016ApJ...823..114N}). Asteroseismology, the study of oscillations in stars, directly constrains the masses of giants. With the increased availability of asteroseismic data, red giants are now more indispensable than ever for stellar age determination, further solidifying their role in understanding the evolution of our universe.

Given the importance of precise age estimations, many tools have been developed to model the correlation between a star's age and its observable characteristics. The astronomy community has produced several variations of models (e.g. MESA Isochrones and Stellar Tracks, MIST, \citealt{2016ApJ...823..102C}; PAdova and tRieste Stellar Evolutionary Code, PARSEC v1.2S/v2.0, \citealt{2015Chen}, \citealt{2022A&A...665A.126N}; Dartmouth Stellar Evolution Program, DSEP, \citealt{2008ApJS..178...89D}; Yonsei-Yale, Y$^2$/YaPSI \citealt{2017ApJ...838..161S}; A Bag of Stellar Tracks and Isochrones, BaSTI-IAC, \citealt{2018Hidalgo}, \citealt{2021Pretrinferni}, \citealt{2022Salaris},  \citealt{2024MNRAS.527.2065P}; Geneva\footnote{\url{https://www.unige.ch/sciences/astro/evolution/recherche/geneva-grids-stellar-evolution-models}},\citealt{2012Ekstrom}, \citealt{2012Georgy}, \citealt{2013Georgy}, \citealt{2019Groh}, \citealt{2021Murphy}, \citealt{2021Eggenberger} \citealt{2022Yusof}, \citealt{2024Sibony}, see footnote for complete list of references; Pisa, \citealt{2011Tognelli}, \citealt{2012Dellomodarme}, \citealt{2013Valle,2013ValleDell}, \citealt{2014Valle, 2014ValleDell}), all of which have their own numerical methods for solving the equations of stellar structure. Researchers then create model grids by applying specific physical assumptions and calibration choices, which can lead to shifts in the correlation between observables and age. Many of these assumptions are about physics that is truly uncertain, including convective overshoot (\citealt{1965Roxburgh}; \citealt{1965Saslaw}; \citealt{2007Claret}; \citealt{CCO}), opacities (\citealt{1996Iglesias}; \citealt{2005Seaton}; \citealt{2016Colgan}; \citealt{opacites}), mixing length (\citealt{1958BohmVitense}; \citealt{2017ApJ...840...17T}; \citealt{2022Cinquegrana}; \citealt{MLT}), nuclear reaction rates (\citealt{2005Weiss}; \citealt{2010Pietrinferni}; \citealt{2023PhLB..84438093D}), equation of state (\citealt{1996Rogers, Rogers2002}; \citealt{EOS}; \citealt{2024MNRAS.527L.179P}), diffusion (\citealt{1917Chapman, 1917bChapman}; \citealt{2002VandenBerg}; \citealt{dotterdiffusion}; \citealt{2022Moedas}), and rotation (\citealt{1989Pinsonneault}; \citealt{2015Gallet}; \citealt{2019ApJ...872..128V}; \citealt{2021Eggenberger}; \citealt{2023A&A...677L...5M}).
 
At this time, there is seldom adequate data to favor one of these models over another in a broad context. However, in individual regimes, researchers can conduct careful examination of individual stars and clusters by comparing isochrone fitting to discern which models are inadequate (\citealt{2020MNRAS.497.3674G}; \citealt{2022AJ....164...34M}; \citealt{2023AJ....165....6S}; \citealt{2023AJ....166...29L}). Such studies highlight the importance of comparing ages from different stellar models to enhance the accuracy of our approximations. Consequently, in age estimations, two sets of uncertainties emerge: one pertaining to observables, which is typically well accounted for, and another concerning theoretical aspects, which can often be overlooked. Recent work has provided a method \citep{2022ApJ...927...31T} to determine realistic uncertainties for solar-type stars on the main sequence, but it is apparent that such techniques are necessary for the stars evolving off the main sequence as they enter the red giant branch. The inconsistencies between models become increasingly evident at this stage due to the diversity in underlying physics of the stellar interior, particularly in choices regarding convection. In this paper, we aim to explore the discrepancies in the red giant branch between four sets of commonly accepted stellar models in order to provide realistic uncertainties in stellar ages. We note that these are not the only available stellar model grids, but were selected for their coverage across the mass and metallicity ranges relevant to this analysis.

\section{Methods}\label{Method}
\subsection{Interpolation}
The main utility used in this paper, {\fontfamily{qcr}\selectfont kiauhoku} \citep{2020ApJ...888...43C, kiauhoku}, is a Python module designed for interpolation between stellar evolutionary tracks by resampling to equivalent evolution phases (EEPs; \citealt{2016Dotter}). To fit models to data, \texttt{kiauhoku} uses the Nelder-Mead optimizer in SciPy (\citealt{Nelder&Mead}; \citealt{2020SciPy-NMeth}) to minimize a user-specified loss function to within a user-specified tolerance \citep{2022ApJ...927...31T}. We employ a mean-squared-error loss with a tolerance of $10^{-6}$.
In the updated version used in this work, we have resampled the models to include more EEPs at the late end of the tracks to improve convergence for the YREC, DSEP, and GARSTEC tracks. However, the MIST {\footnote{\label{sharednote}\url{https://waps.cfa.harvard.edu/MIST/}}} website only provides the downsampled EEP tracks{\footnote{\url{https://waps.cfa.harvard.edu/MIST/model_grids.html\#eeps}}} and access to the full-resolution grids is necessary to resample without introducing numerical error. As a result, we selected the non-rotating, theoretical tracks at the default resolution provided by MIST. Our interpolation process uses mass, metallicity, surface gravity, luminosity, and/or effective temperature to fit for ages. To simulate the analysis of different observational methods, we consider stars with masses ranging from $0.6$ to $2.0\,\msun$, metallicities between $\mathrm{[Fe/H]}=-1.0$ and $+0.5$, surface gravities $\log(g) = 0.0$ to $3.5$, luminosities spanning $10$ to $2000\,L_\odot$, and effective temperatures between $3000$ and $5500$\,K. 

\subsection{Grid Selection}
We make use of widely used model grids to examine the theoretical uncertainties stemming from various choices in model physics. We sample four model grids: Yale Rotation Evolution Code (YREC; \citealt{2022ApJ...927...31T}; \citealt{1989Pinsonneault}), MESA Isochrones and Stellar Tracks (MIST; \citealt{2016ApJ...823..102C}; \citealt{2011Paxton}), Dartmouth Stellar Evolution Program (DSEP; \citealt{2008ApJS..178...89D}; \citealt{2001Chaboyer}), and Garching Stellar Evolution Code (GARSTEC; \citealt{2013MNRAS.429.3645S}; \citealt{2008Ap&SS.316...99W}). Table 1 from \citet{2022ApJ...927...31T} outlines the model physics for each case. For the reader's convenience, this table has been replicated below as Table \ref{Table:physics}. These models have notable differences ranging from solar abundances chosen to the internal processes of stars (i.e: diffusion, convection overshoot etc.). The following sections will briefly go over input physics relevant to the location of the RGB. 

\begin{deluxetable*}{p{3cm}p{3cm}p{3cm}p{3cm}p{3cm}} 
\tabletypesize{\scriptsize}  
\tablewidth{0.85\textwidth}
\tablecaption{Summary of Input Physics for Each Model Grid\textsuperscript{a} \label{Table:physics}}
\tablehead{
    \colhead{Parameter} & \colhead{YREC} & \colhead{MIST} & \colhead{DSEP} & \colhead{GARSTEC}}
\startdata
Reference&\citet{2022ApJ...927...31T}&\citet{2016ApJ...823..102C}&\citet{2008ApJS..178...89D}&\citet{2013MNRAS.429.3645S}\\
Atmosphere & Grey & \citet{1993Kurucz} & PHOENIX \citep{1999Ha, 1999Hb} & Grey \\
Convective Overshoot & Step: 0.16H$_{\mathrm{p}}$ & Diffusive: 0.0160 (core) and 0.0174 (env) & Step: 0.2H$_{\mathrm{p}}$ & Diffusive: 0.02 \\
Diffusion & Yes & Main Sequence only & Modified & Yes \\
Equation of State & OPAL {+} SCVH\textsuperscript{b,c} & OPAL {+} SCVH {+} MacDonald {+} HELM {+} PC\textsuperscript{c,d,e} & Ideal Gas with \citet{ChaboyerKim1995} + \citet{Irwin2004} & \citet{Irwin2004} \\
High-Temperature Opacities & OPAL\textsuperscript{g} & OPAL\textsuperscript{f,g} & OPAL\textsuperscript{g} & OPAL\textsuperscript{g} \\
Low-Temperature Opacities & \citet{Ferguson2005} & \citet{Ferguson2005} & \citet{Ferguson2005} & \citet{Ferguson2005} \\
Mixing Length & \citet{2017ApJ...840...17T} & 1.82 & 1.938 & 1.811 \\
Nuclear Reaction Rates & \citet{Adelberger2011} & \citet{Cyburt2010} & \citet{Adelberger1998} {+} \citet{2004Imbriani} {+} \citet{Kunz2002} {+} \citet{Angulo1999} & \citet{Adelberger1998} {+} \citet{Angulo1999} \\
Rotation & \citet{TayarPinsonneault2018} & None & None & None \\
Weak Screening & \citet{Salpeter1954} & \citet{AlastueyJancovici1978} & \citet{Salpeter1954} {+} \citet{Graboske1973} & \citet{Salpeter1954} \\
Mixture and Solar $\mathrm{Z/X}$ & \citet{1998GS} & \citet{2009Asplund} (protosolar) & \citet{1998GS} & \citet{1998GS} \\
Solar X & 0.709452 & 0.7154 & 0.7071 & 0.7090 \\
Solar Y & 0.2725693 & 0.2703 & 0.27402 & 0.2716 \\
Solar Z & 0.0179492 & 0.0142 & 0.01885 & 0.0193 \\
$\Delta\mathrm{Y/}\Delta\mathrm{Z}$ & 1.3426 & 1.5 & 1.5327 & 1.194 \\
Surface $\mathrm{(Z/X)}_\odot$ & 0.0253 & 0.0173 & 0.0229 & 0.0245
\enddata
{\scriptsize
\tablenotetext{a}{Table reproduced from \citet{2022ApJ...927...31T} for reference.}
{\textsuperscript{b}~\citet{1996Rogers}; 
\textsuperscript{c}~\citet{RogersNayfonov2002}; 
\textsuperscript{d}~\citet{Saumon1995}; 
\textsuperscript{e}~\citet{2012MacdonaldMullan}; 
\textsuperscript{f}~\citet{1993Iglesias}; 
\textsuperscript{g}~\citet{1996Iglesias}.}
}
\end{deluxetable*}

\subsubsection{YREC}
The YREC models used here, generated using the YREC code \citep{1989Pinsonneault}, were first introduced in \citet{2022ApJ...927...31T}. These tracks were calibrated to match the observed properties of red giants and implemented a metallicity dependent mixing length (\citealt{1958BohmVitense}; \citealt{2017ApJ...840...17T}). This version of YREC adopts a gray atmosphere and the \citet{1998GS} abundances. Convective overshoot, diffusion, and rotational evolution are included in the models, but rotational mixing is not.

\subsubsection{MIST}
The MIST evolutionary tracks \citep{2016ApJ...823..102C} were created using the MESA Stellar Evolution Code \citep{2011Paxton, Paxton2013, Paxton2015, Paxton2018, Paxton2019} and are available online. These models \footref{sharednote} were calibrated to reproduce helioseismic data and surface properties of the Sun by varying the composition, mixing length parameter ($\alpha_{MLT}$), and convective overshoot in the envelope as it evolves from the pre-main sequence to 4.57 Gyrs. MIST adopts the \citet{1993Kurucz} boundary conditions with \citet{2009Asplund} abundances and a mixing length parameter $\alpha_{MLT} = 1.82$. The models apply diffusion only on the main sequence and include overshoot for both convective cores and envelopes. In this study, we use the non-rotating version of the models; however, rotating models are available online and models with magnetic braking are described in \citet{2021Gossage}. These models have been tested across both low-mass and high-mass regimes, using observational data from globular clusters, open clusters, and the pre-main-sequence quadruple system LkCa 3 \citep[see][and references therein]{2016ApJ...823..102C}.

\subsubsection{DSEP}
The DSEP models used here were first presented by \citet{2008ApJS..178...89D} and generated using the Dartmouth Stellar Evolution Program \citep{2001Chaboyer}. DSEP adopts PHOENIX (\citealt{1999Ha};\citealt{1999Hb}) atmosphere conditions for the low temperature regime (T $< 10,000$ K), \citet{1993Kurucz} for the high temperature regime (T $>10,000$ K), and \citet{1998GS} abundances. The models include atomic diffusion, including gravitational settling, which is applied in the stellar interior and partially supressed in the outermost 0.1 $M_\odot$. DSEP handles convection with the standard mixing length theory and uses a solar-calibrated $\alpha_{MLT} = 1.938$. Convective overshoot is included, but rotation is not. These DSEP models have been validated through comparisons of data of globular and open clusters \citep[see][and references therein]{2008ApJS..178...89D}.

\subsubsection{GARSTEC}
 GARSTEC is a stellar evolution code originally presented in \citet{2008Ap&SS.316...99W}, and has been used to produce several model grids. In this work we use the grid presented in  \citet{2013MNRAS.429.3645S}, which we refer to as the GARSTEC grid for simplicity. While models created with the GARSTEC code are not distributed as a formal model library, we include this version to diversify the range of model physics represented in our analysis and have made the tracks available online.{\footnote{\url{https://zenodo.org/records/14908017}}}. The models use a gray atmosphere and \citet{1998GS} abundances. Convection is treated using a solar-calibrated mixing length parameter $\alpha_{MLT} = 1.811$. Diffusive convective overshoot is modeled following \citet{1996Frey}, but modified to decrease quadratically with the ratio of convective core size and the pressure scale height at the boundary. Rotation is not included in the models. Unlike other model sets used in this study, this particular GARSTEC grid was created for the development of a Bayesian analysis to determine the masses and ages of stars using non-LTE spectroscopic parameters \citep{2013MNRAS.429.3645S}.

\section{Demonstrative Examples}\label{DemoEx}

\subsection{Motivation}
\begin{figure}
    \centering
    \includegraphics[scale=.6]{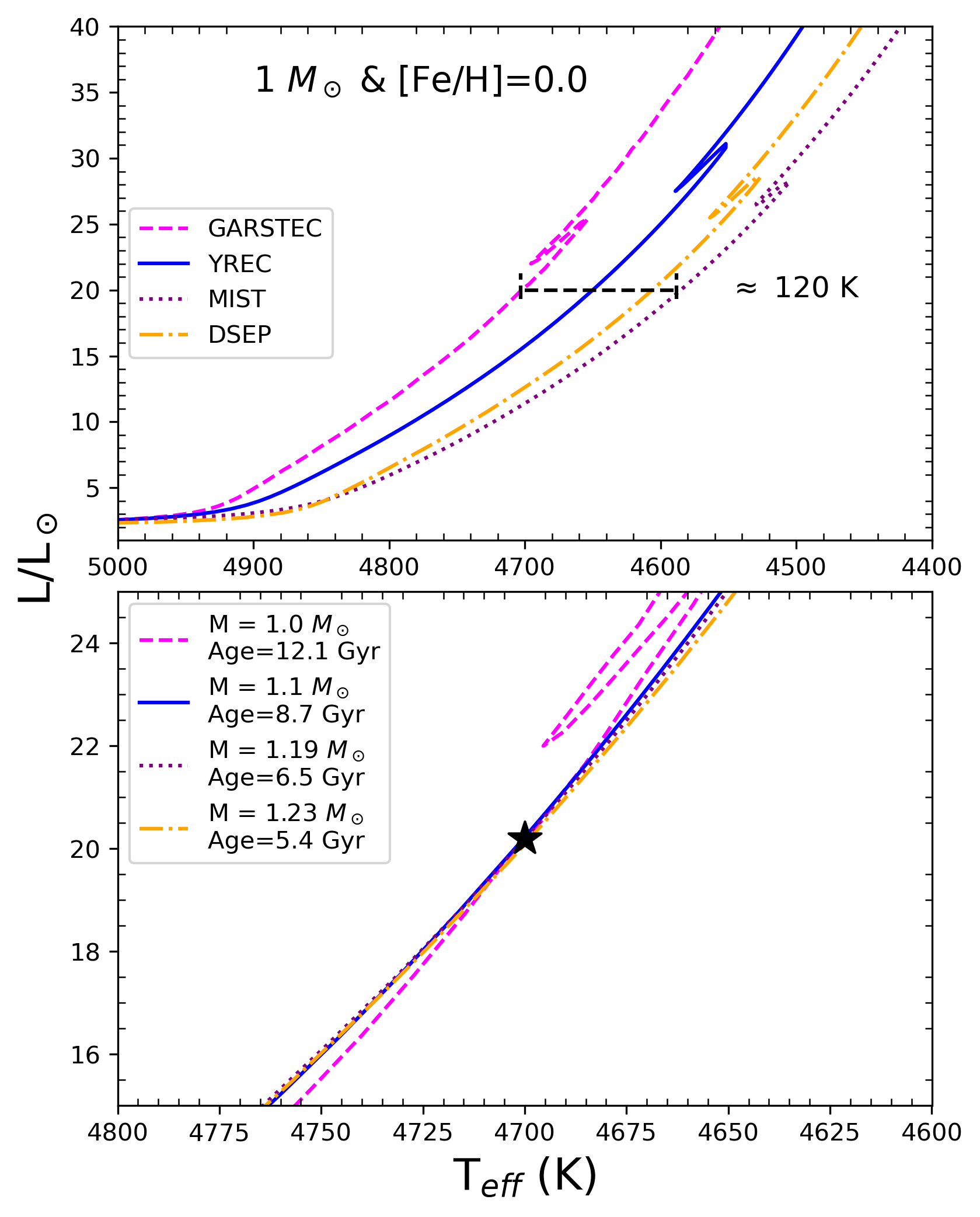}
    \caption{\textit{Top}: Luminosity vs. Effective Temperature for a $1\msun$ star at [Fe/H]=0.0, with model tracks from GARSTEC (pink), YREC (blue), MIST (purple), and DSEP (orange). The black dashed line shows a temperature spread of $\approx 120 $K at $L/L_\odot = 20$. \textit{Bottom}: Luminosity vs. Effective Temperature at an intersection point of the model tracks (fixed $L/L_\odot = 20$, [Fe/H]=0.0, and $T_{eff} = 4700$ K), showing a mass difference of 23\%. This mass difference corresponds to an age difference of 55\%.}
    \label{figure:1}
\end{figure}

To make clear the necessity of assessing the theoretical uncertainties, we show in the top panel of Figure \ref{figure:1} the four different evolutionary tracks for a $1\msun$ star, showing how luminosity varies with effective temperature at solar metallicity. Examining this case, we observe that the tracks do not align, resulting in a temperature difference of $\approx 120~K$ at a luminosity of $20L_\odot$. Within a single model grid, a $0.1\msun$ change in mass typically results in a temperature shift of around 30 K, suggesting that a $\approx 120~K$ offset could be a significant change in mass, though this mass-temperature sensitivity differs slightly between models.

When we constrain our comparison purely to observable parameters, fixing luminosity at $20L_\odot$, solar metallicity [Fe/H]=0.0, and temperature at 4700 K, we find that the inferred masses differ by approximately 23\% across model grids. This discrepancy in mass leads to corresponding grid age estimations that differ by about 55\%. 

The predicted effective temperature scale for red giant stars is known to be highly sensitive to a range of physical inputs, including the treatment of convective mixing, outer boundary conditions, low-temperature opacities, and the adopted heavy element distribution (\citealt{2002Salaris}; \citealt{2008Vandenberg}; \citealt{Choi18}; \citealt{2018Salaris}; \citealt{2024Valle}; \citealt{2024Creevey}). While many model grids are calibrated to reproduce the Sun’s properties at solar metallicity, this constraint does not fully eliminate differences in the predicted temperature scale, especially along the RGB.

These types of differences are not consistent across models. It is likely that while the models can agree in some regions, they can also disagree in others. To explore these inconsistencies further, we analyzed commonly accessible observable parameters, including asteroseismic data from Kepler and APOGEE, spectroscopic data from APOGEE, and photometric data from Gaia, to quantify variations in estimated ages across models.

\subsection{Asteroseismology}\label{Seis}
\begin{figure*}
    \centering
    \includegraphics[scale=.45]{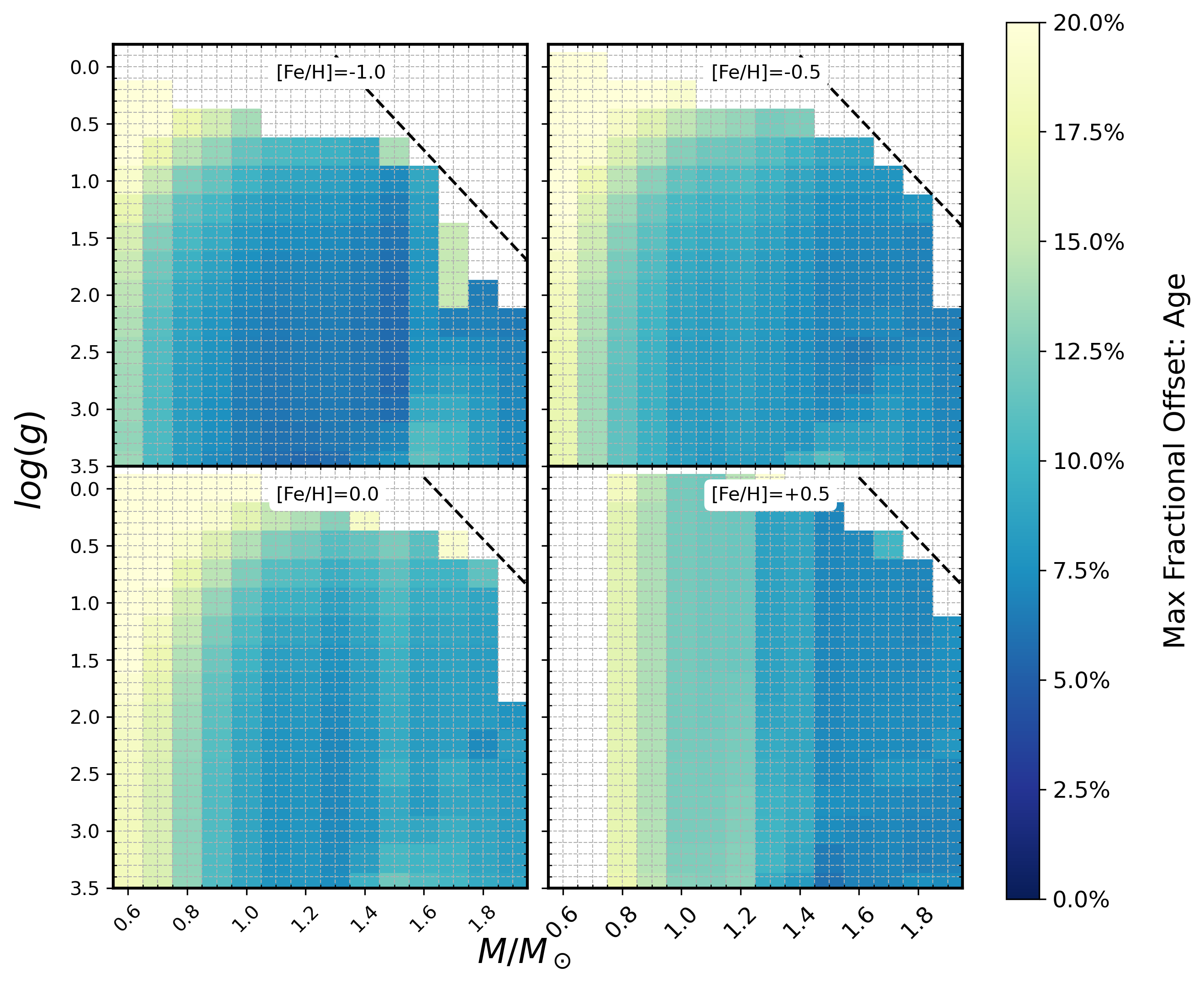}
    \caption{Mass vs. $\log(g)$: Maximum fractional offset in age between model grids for mass range $0.6-1.9 \msun$ and surface gravity range $\mathrm{log(g)}=0.0-3.5$ for metallicities $\mathrm{[Fe/H]} = [-1.0, -0.5, 0.0, +0.5]$. The black dotted line represents the tip of the giant branch for each mass track. The mean offset in age between models for each metallicity is $9.4\%$, $11.1\%$, $12.0\%$, and $10.4\%$, respectively. This indicates that age estimations are in good agreement with each other when mass is known.}
    \label{figure:2}
\end{figure*}

We begin by considering a favorable test case in which a hypothetical star has been analyzed using asteroseismology and spectroscopy, providing known values for mass, metallicity, and surface gravity. We examined stars within the mass range $0.6-2.0 \msun$ and surface gravity range $\mathrm{log(g)}=0.0-3.5$ for metallicities $\mathrm{[Fe/H]} = [-1.0, -0.5, 0.0, +0.5]$. For each point in Figure \ref{figure:2}, we show the maximum fractional offset in age. Figure \ref{figure:2} shows that for stars of known mass, the models are in general agreement with the highest differences being at low mass and low metallicity, reaching an age difference of $30\%$. The vertical banding present in the figure results from the minimal variation in age for stars of a particular mass as they decrease in surface gravity (i.e., ascend the RGB). The mean offset for each metallicity regime is $9\%$, $11\%$, $12\%$, and $10\%$ for $\mathrm{[Fe/H]} = [-1.0, -0.5, 0.0, +0.5]$, respectively, consistent with previous grid-to-grid offsets found in \citet{2025Pinsonneault}. This indicates that for stars with mass and metallicity values close to solar, model choice is less critical because all grids will give a similar age estimate, with offsets less than $12\%$. For stars in the higher metallicity range ($\mathrm{[Fe/H] = +0.5}$), offsets reach $ \approx 15\%$ at a mass of $1.25\msun$, increasing with decreasing mass. 

By comparing these model-dependent offsets to typical observational uncertainties, we can identify the dominant source of age uncertainty. When mass, metallicity, and surface gravity are known, the theoretical grids remain the primary source of uncertainty when considering observational errors of 5.5\%, 0.058 dex, and 0.005 dex for mass, metallicity, and surface gravity, respectively (\citealt{2025Pinsonneault}). For instance, for a $1 M_\odot$ star with [Fe/H] $=0.0$ and $\log(g)=2.0$, these observational uncertainties result in an age offset of approximately $10\%$, slightly smaller than the theoretical uncertainty.

\subsection{Spectroscopy}\label{sec:spec}
\begin{figure}
    \centering
    \includegraphics[scale=.45]{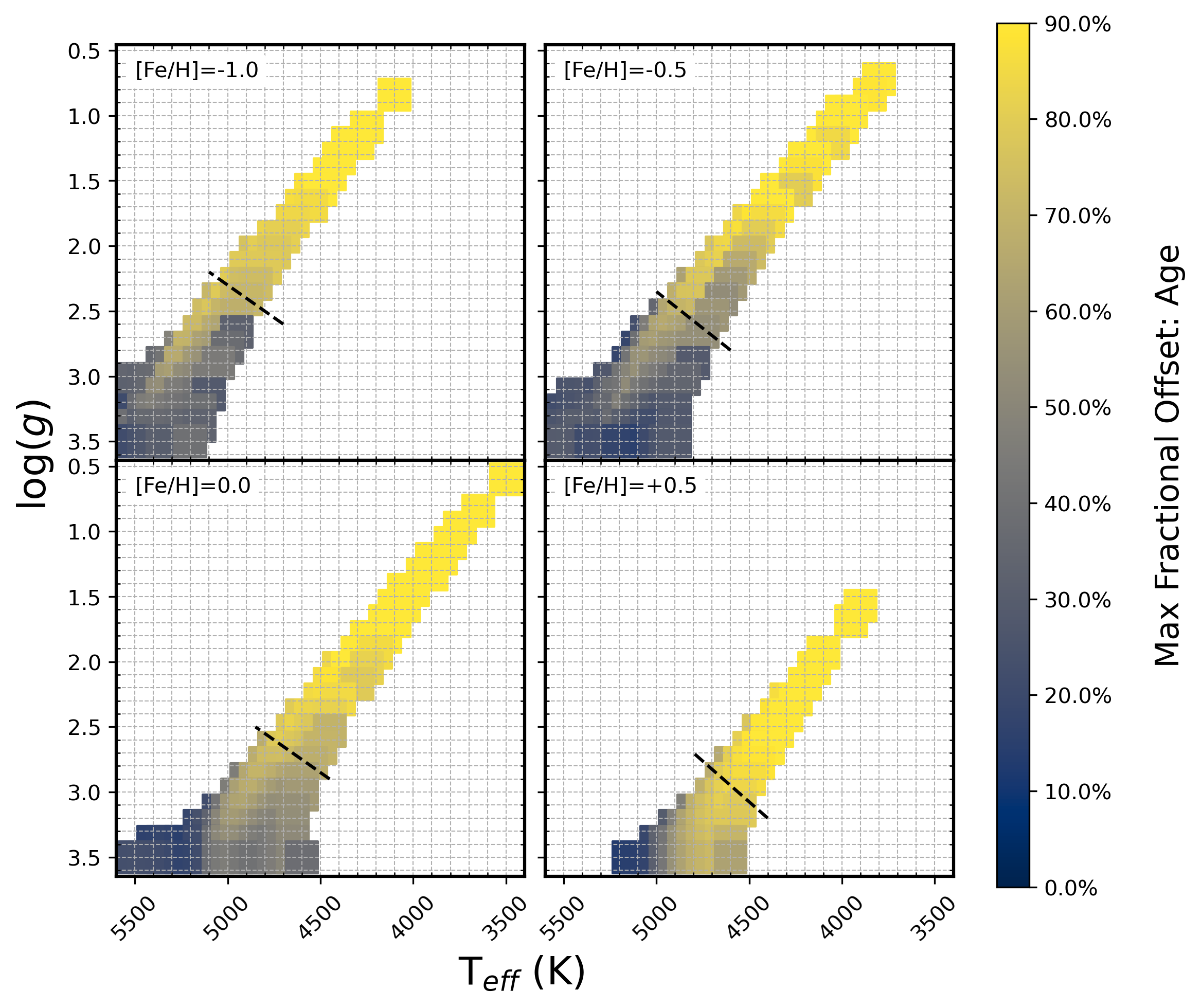}
    \caption{Effective Temperature vs. Surface gravity: Maximum fractional offset in age between model grids for temperature range $\mathrm{T}_{eff} = 3000 - 5500 K$ and surface gravity range $\mathrm{log(g)}=0.0-3.5$ for metallicities $\mathrm{[Fe/H]}= [-1.0, -0.5, 0.0, +0.5]$. The black dashed line indicates the average position of the RGB bump across all grids. The mean age offset reaches up to $60\%$ before the bump and increases to $80\%$ beyond the bump for each metallicity. The mean age offsets for each metallicity are 57\%, 51\%, 62\%, and 69\%, respectively.}
    \label{figure:3}
\end{figure}

Given that asteroseismology is relatively rare while millions of stars have spectroscopic data, our second test case involves a hypothetical star analyzed using high-resolution spectroscopy, providing known values for surface gravity, metallicity, and effective temperature. We note that the current methods for inferring age from spectroscopy often incorporate a combination of asteroseismic data, photometric data, or machine learning (\citealt{2019Das};  \citealt{2023Anders}; \citealt{2024Zhang}). However, we believe that it is best to examine this case in its simplest form, using only spectroscopic parameters. We examined stars within the surface gravity range $\mathrm{log(g)} = 0.0-3.5$ and the temperature range $\mathrm{T}_{eff} = 3000 - 5500 K$ for metallicities $\mathrm{[Fe/H]}= [-1.0, -0.5, 0.0, +0.5]$. Our results in Figure \ref{figure:3} indicate that the maximum fractional offset in age between grids increases as we climb the RGB. The differences between grids reach $\approx 60\%$ at $\mathrm{log(g)} \approx 2.5$ ($\approx 3.0$ for high metallicity) where the RGB bump is expected. After the bump, the offsets in age reach $\approx 90\%$, which poses significant challenges not only for studies of distant galaxies where we rely on the brightest stars for crucial information about stellar populations and star formation histories, but also for examining faraway regions of our own galaxy, such as the halo.

While our analysis focuses on the spectroscopy-only case, even studies that incorporate photometry to estimate mass report considerable age uncertainties. For example, \citet{2016Feuillet} used PARSEC models and found an age uncertainty of 0.15 dex (~41\%) when mass was derived from photometric and spectroscopic parameters, with typical mass uncertainties of ~0.3 dex. In that study, a mock sample was used to test whether the same mass could be recovered from isochrone fitting with just observables, and the resulting mass estimates agreed with the calculated mass from observables to within ~38\%, consistent with expectations from observational uncertainties. Based on our findings in Section \ref{Seis}, even though the PARSEC models used in that study are not directly compared here, we can reasonably infer that model-dependent differences would contribute an additional ~10\% uncertainty in age if multiple stellar evolution models were considered. However, our spectroscopy-only analysis showcases that in the absence of a mass constraint, each model must infer mass independently from the observable parameters, leading to differing mass estimates and, in turn, significantly larger discrepancies in age, even before accounting for uncertainties in the observables. This highlights the critical role of mass as a unifying constraint for achieving consistent age determinations across models.

\begin{figure}
    \centering
    \includegraphics[scale=.45]{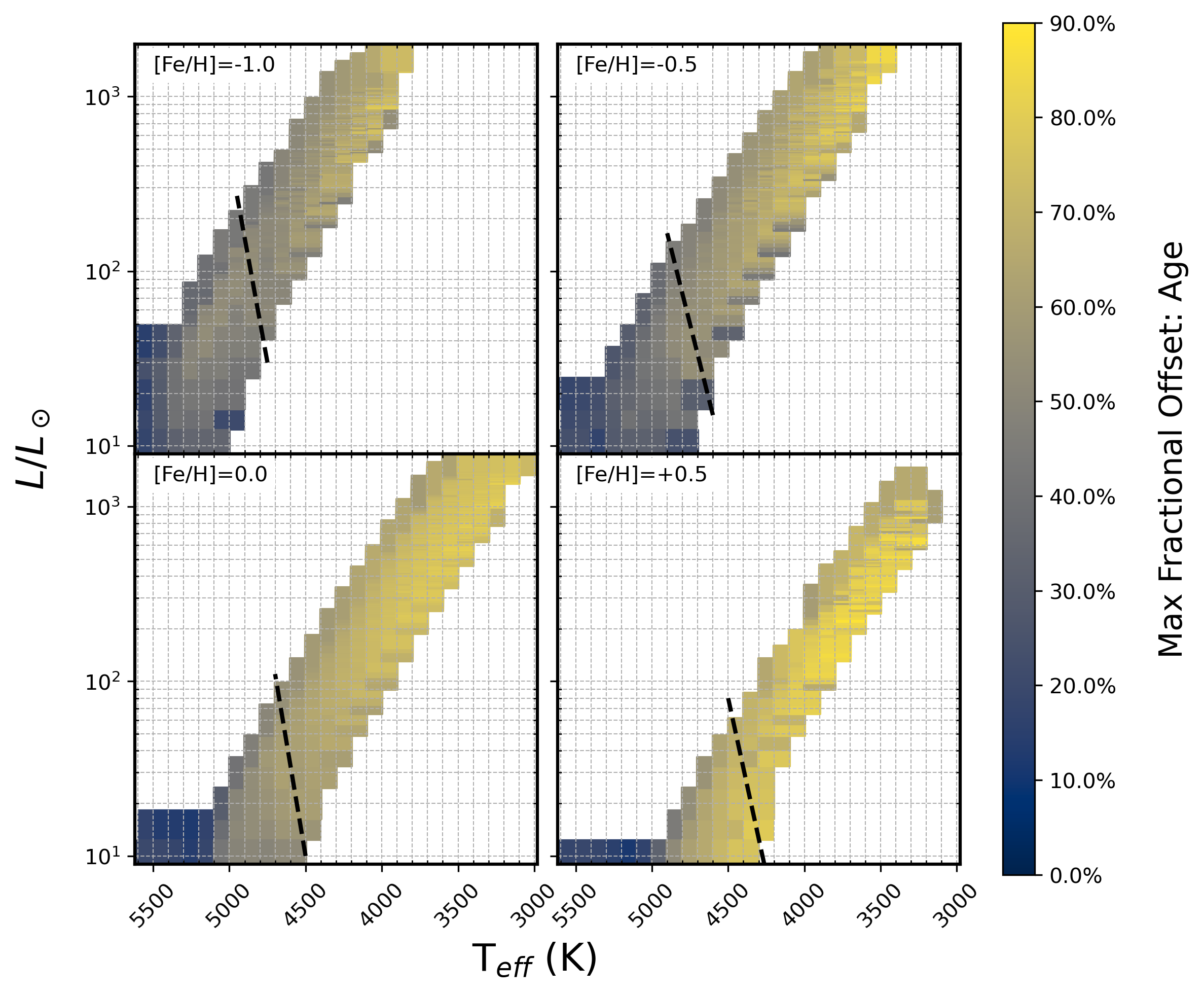}
    \caption{Effective Temperature vs. Luminosity: Maximum fractional offset in age between model grids for temperature range $\mathrm{T}_{eff} = 3000 - 5500$ K and luminosity range $L/L_\odot = 10-2000$ for metallicities $\mathrm{[Fe/H]}= [-1.0, -0.5, 0.0, +0.5]$. The black dashed line indicates the average position of the RGB bump across all grids. The mean age offsets for each metallicity are 57\%, 64\%, 72\%, and 74\%, respectively.}
    \label{figure:4}
\end{figure}

\subsection{Gaia Photometry}
Considering that missions like \textit{Gaia} \citep{2023Gaia}, offer extensive data on over 33 million stars, our last test case is a star analyzed with photometry and astrometry, which has been used to infer luminosity, metallicity, and effective temperature. We examine stars within a luminosity range of $L/L_\odot = 10-2000$,  temperature range of $\mathrm{T}_{eff} = 3000 - 5500$ K, and the same metallicity regimes as previously used. Figure \ref{figure:4} illustrates a similar trend to that observed in Spectroscopy (Section \ref{sec:spec}) where the age differences increase with luminosity reaching $\approx 60\%$ before the RGB bump and $\approx 80\%$ after. We are not the first to notice that traditional photometric methods can produce unreliable ages for some giants. For example, a previous study reported age estimates with $20\%$ accuracy for MS and sub-giant stars but was unable to determine reliable ages for red giants, as the PARSEC models predicted RGB temperatures that were too high compared to observations \citep{2019Howes}. This limitation illustrates the risks of relying on one evolutionary grid, and our analysis further demonstrates how uncertainties increase when considering multiple models.  

\section{Applications to Data}
\subsection{APOKASC-3}\label{Apo}
As a final demonstration, we illustrate how these theoretical uncertainties would apply to real data using measurements from the APOKASC-3 catalog, which includes a substantial sample of 15,809 evolved stars. These stars have been analyzed with APOGEE, as part of the Sloan Digital Sky Survey (SDSS; \citealt{2000sdss}, SDSS-III; \citealt{2011sdss2E}, SDSS-IV; \citealt{2017sdss3B}) spectroscopic parameters and Kepler asteroseismic data using ten independent techniques \citep{2025Pinsonneault}. Of this sample, 12,448 stars have precise asteroseismic measurements of mass, radius, and age. The estimated ages were computed using a variant of the GARSTEC grids, which have been modified to include the solar mixture scale of \citet{Grevesse1993InOA} and helium abundance $\Delta Y/\Delta Z = 1.15$. Convection is treated using a mixing length prescription by \citet{1968Cox&Giuli} and uses a solar-calibrated mixing length parameter $\alpha_{MLT}=2.012$ \citep{2025Pinsonneault}. In our sample cuts, we excluded stars outside of our specified mass and metallicity ranges, as well as red clump stars. For this analysis, we did not exclude potential binaries or mass transfer products, so we anticipate that these stars will provide biased age estimates. We recommend that researchers who use these data for Galactic Archaeology carefully evaluate the impact of binaries on their findings (\citealt{2023Cehula}; \citealt{2023Bufanda}; \citealt{2024Patton}). After applying our selection criteria, our final sample consists of 4,339 giants. Metallicities were adjusted using the Salaris correction \citep{1993Salaris} to account for $\alpha$-enhancement, as a more precise estimate of the heavy element distribution is crucial for accurate temperature estimates \citep{2018Salaris}. We then applied the same procedure as in the earlier examples to emphasize how the choice in observational parameters affects the model-estimated ages. 

\subsubsection{APOKASC-3: Asteroseismic Parameters}\label{aposeis}
\begin{figure}[hbt]
    \centering
    \includegraphics[scale=.4]{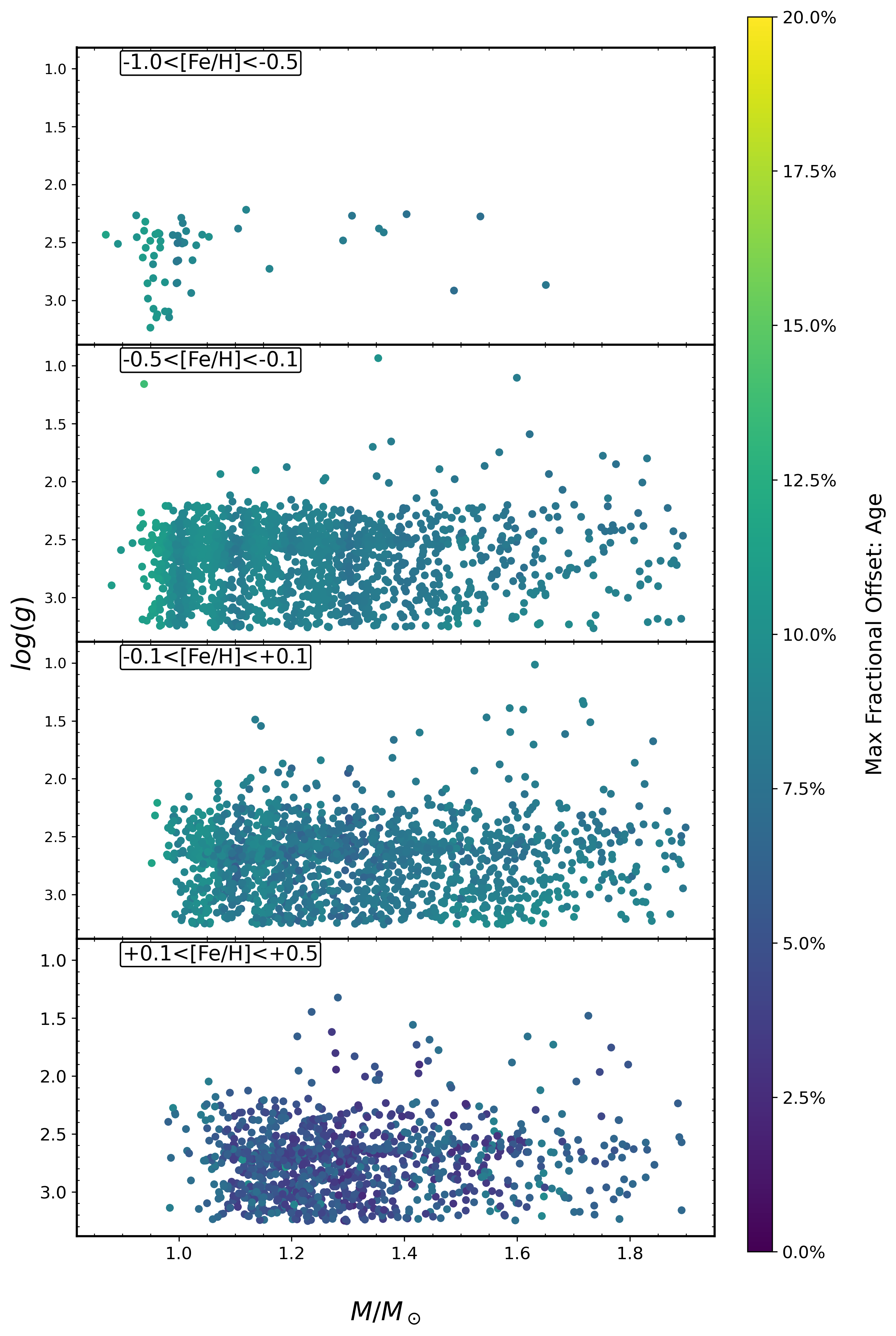}
    \caption{Mass vs surface gravity for RGs in the APOKASC-3 sample. We show the maximum fractional offset in age between the model grids organized by metallicity. When interpolating with mass as a known value, this sample is consistent with our previous demonstrations, where theoretical offsets between model grids are less than 10\%. Starting from the top panel, the corresponding averages in age offset are 9\%, 9\%, 8\%, and 5\%.}
    \label{figure:5}
\end{figure}

We began by replicating our first scenario, by interpolating ages based on asteroseismic mass, metallicity, and surface gravity. We excluded stars with estimated ages exceeding 15 Gyrs in at least 3 out of the 4 grids from both the averages and the figures, as these are likely to be classified as potential binary interaction products, which would typically be removed from age distributions (\citealt{2005Jorgensen}; \citealt{2024Horta}). In Figure \ref{figure:5}, we show the maximum fractional offset in age between the grids, organized by metallicity ranges: $-1.0<[\mathrm{Fe/H}]<-0.5$, $-0.5<[\mathrm{Fe/H}]<-0.1$, $-0.1<[\mathrm{Fe/H}]<+0.1$, and $+0.1<[\mathrm{Fe/H}]<+0.5$. Consistent with previous findings, the average age offsets between models for each regime are 9\%, 9\%, 8\%, and 5\%, with small trends as a function of mass, metallicity, and distance from the nearest model \citep{2025Li&Joyce}. Since the APOKASC-3 catalog provides age estimates with a median fractional uncertainty of 11.1\%, incorporating both observational and theoretical uncertainties, its careful approach offers a more comprehensive treatment of factors affecting age inferences, leading to more consistent results. This allows us to assess how well each age estimate from the grids aligns with those from APOKASC-3. To provide more detailed insight, we break each of these plots into the grid-by-grid version presented in the appendix \ref{sec:appendix}. Figure \ref{fig:ex1} shows the offset between the APOKASC-3 catalog's estimated age and the corresponding grid's estimated age. The average offsets are 5\%, 5\%, 4\%, and 4\% for YREC, GARSTEC, DSEP, and MIST, respectively. This is consistent with our results from Section \ref{Seis}, that model choice is less important for red giants with accurate asteroseismic parameters. 

\begin{figure}[hbt]
    \centering
    \includegraphics[scale=.35]{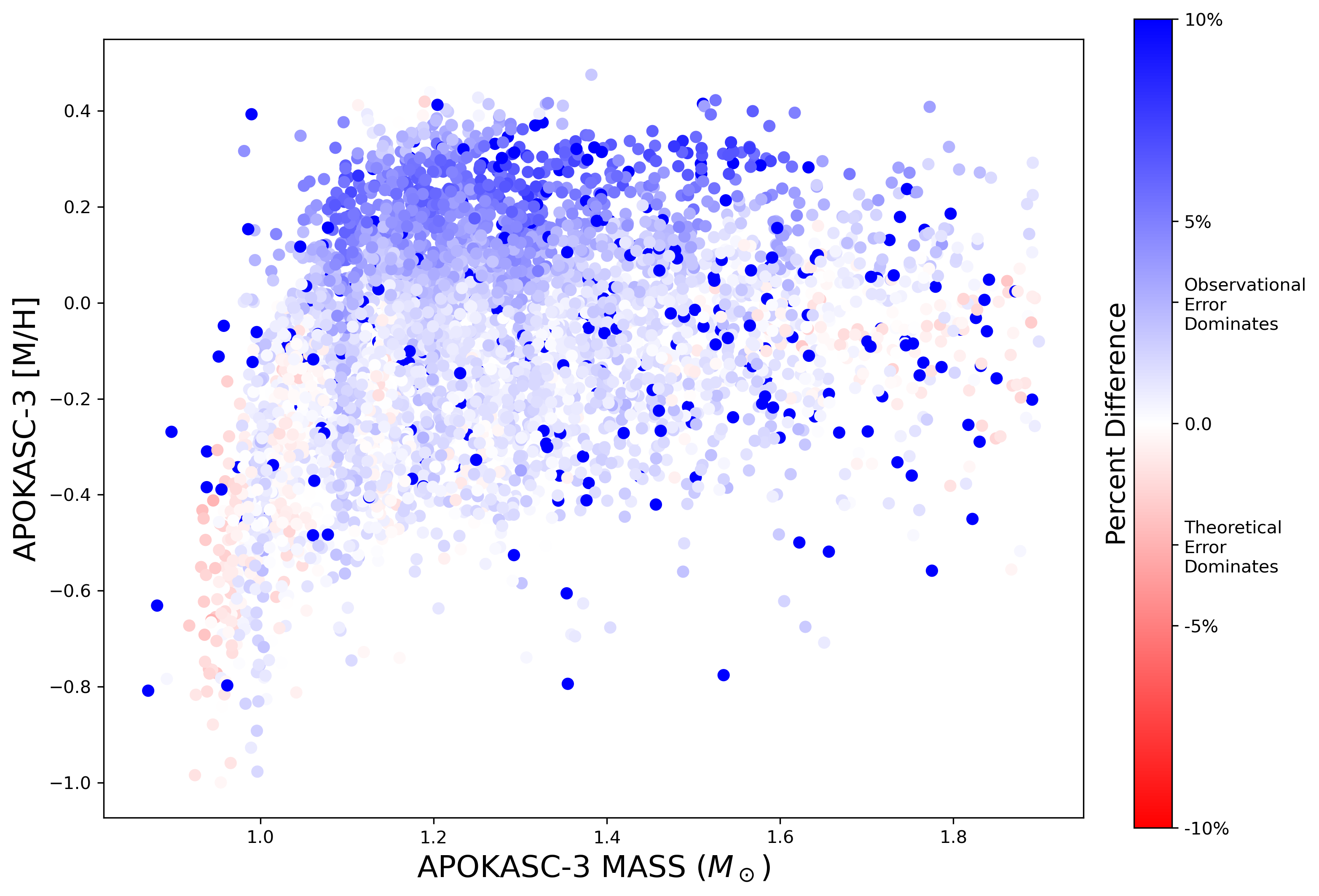}
    \caption{Comparison of observational error to theoretical error resulting from model choice for our asteroseismic interpolation of the APOKASC-3 sample. Our results indicate that observational error dominates over theoretical error for almost all giants, with minor deviations in the low mass, low metallicity regime. }
    \label{seisobserr}
\end{figure}

Next, we explored whether the age offsets between grids exceeded the observational uncertainties. The observational uncertainties were estimated using a Monte Carlo (MC) simulation with 100 iterations, incorporating the uncertainties provided in the catalog and defined as the average across all grids of the 50th percentile of the MC distribution. An example of the Monte Carlo results is included in the appendix (\ref{fig:mcexmass} - \ref{fig:mcexgaia}). Our results, shown in Figure \ref{seisobserr}, indicate that the observational uncertainty outweighed the theoretical uncertainty resulting from model choice for most giants, with minor deviations in the low mass, low metallicity regime, illustrated by using the percent difference between observational and theoretical uncertainty. These findings further solidify the need for incorporating mass in grid age interpolation to obtain more accurate age estimates for stars along the RGB.

\subsubsection{APOKASC-3: Spectroscopic Parameters}
\begin{figure}[hbt]
    \centering
    \includegraphics[scale=.4]{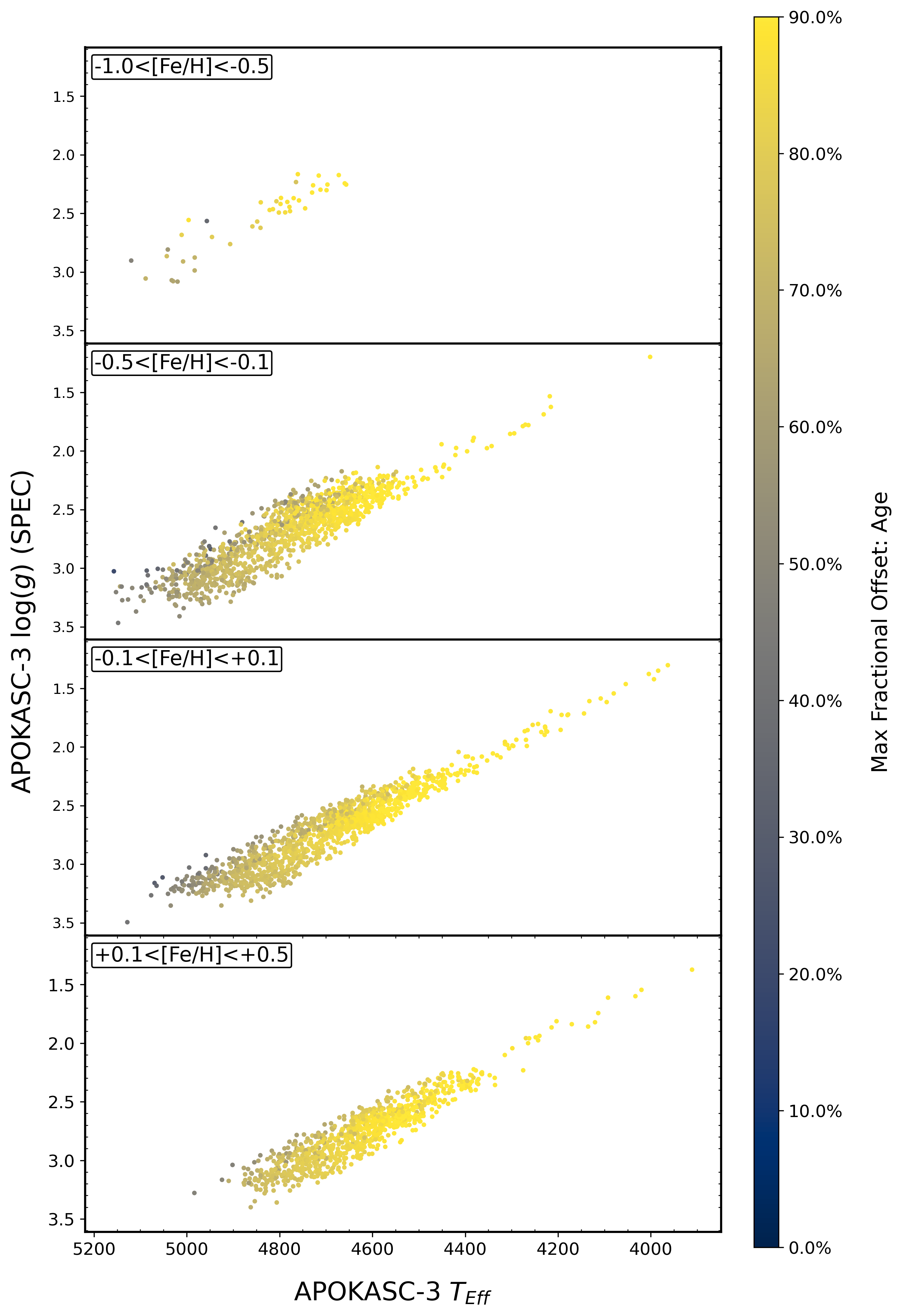}
    \caption{Maximum fractional offset between grids using spectroscopic information from APOKASC-3. We found significant mean offsets across all metallicity regimes to be 80\%, 77\%, 83\%, and 80\% for metallicity regimes $-1.0<[\mathrm{Fe/H}]<-0.5$, $-0.5<[\mathrm{Fe/H}]<-0.1$, $-0.1<[\mathrm{Fe/H}]<+0.1$, and $+0.1<[\mathrm{Fe/H}]<+0.5$, respectively. Our results indicate that offsets in age increase with decreasing surface gravity.}
    \label{specoffset}
\end{figure}

As a next step, we interpolated grid age estimates using the APOKASC-3 spectroscopic surface gravity, metallicity, and APOGEE effective temperature. Figure \ref{specoffset} shows a dramatic increase in age differences when we change to using spectroscopic parameters. We find the average age offsets between grids to be 80\%, 77\%, 83\%, and 80\% for metallicity regimes $-1.0<[\mathrm{Fe/H}]<-0.5$, $-0.5<[\mathrm{Fe/H}]<-0.1$, $-0.1<[\mathrm{Fe/H}]<+0.1$, and $+0.1<[\mathrm{Fe/H}]<+0.5$, respectively. These offsets are even larger than those seen in our  demonstrative example in Figure \ref{figure:3}. However, this still shows that differences in inferred age increase with decreasing surface gravity (i.e., higher up the giant branch). Figure \ref{fig:ex2} shows the offset between the APOKASC-3 catalog's estimated age using asteroseismic results and the corresponding grid's estimated age using only spectroscopic information. The average offsets are 37\%, 100\%, 45\%, and 67\% for YREC, GARSTEC, DSEP, and MIST, respectively. As a reminder to the reader, the age offsets shown here compare two different variants of the GARSTEC models: the version used in our analysis (based on \citealt{2013MNRAS.429.3645S}) and the modified version used in the APOKASC-3 catalog \citep{2025Pinsonneault}. The large offset in age reflects the differences in the input physics between these grids (see Section \ref{Apo}). We also show the offset in mass between the APOKASC-3 stellar masses and the grid estimated masses in Figure \ref{fig:ex3}, averaging at 12\%, 15\%, 18\%, and 43\%. The GARSTEC model grids tend to produce older ages due to its generally hotter temperature scale, which leads to lower inferred masses. As a result, our original condition, excluding stars with estimated ages exceeding 15 Gyrs in at least 3 out of 4 grids, does not work in GARSTEC's favor, as it retains a larger fraction of older age estimates. However, if we remove all stars that GARSTEC estimates an age that exceeds 15 Gyrs, the mean offset decreases to $66\%$ in age and $13\%$ in mass, compared to the APOKASC-3 estimates. Regardless of the selection criteria, these results highlight that age estimates relying solely on spectroscopic data are much less reliable than those that incorporate a known mass.

\begin{figure}[hbt]
    \centering
    \includegraphics[scale=.3]{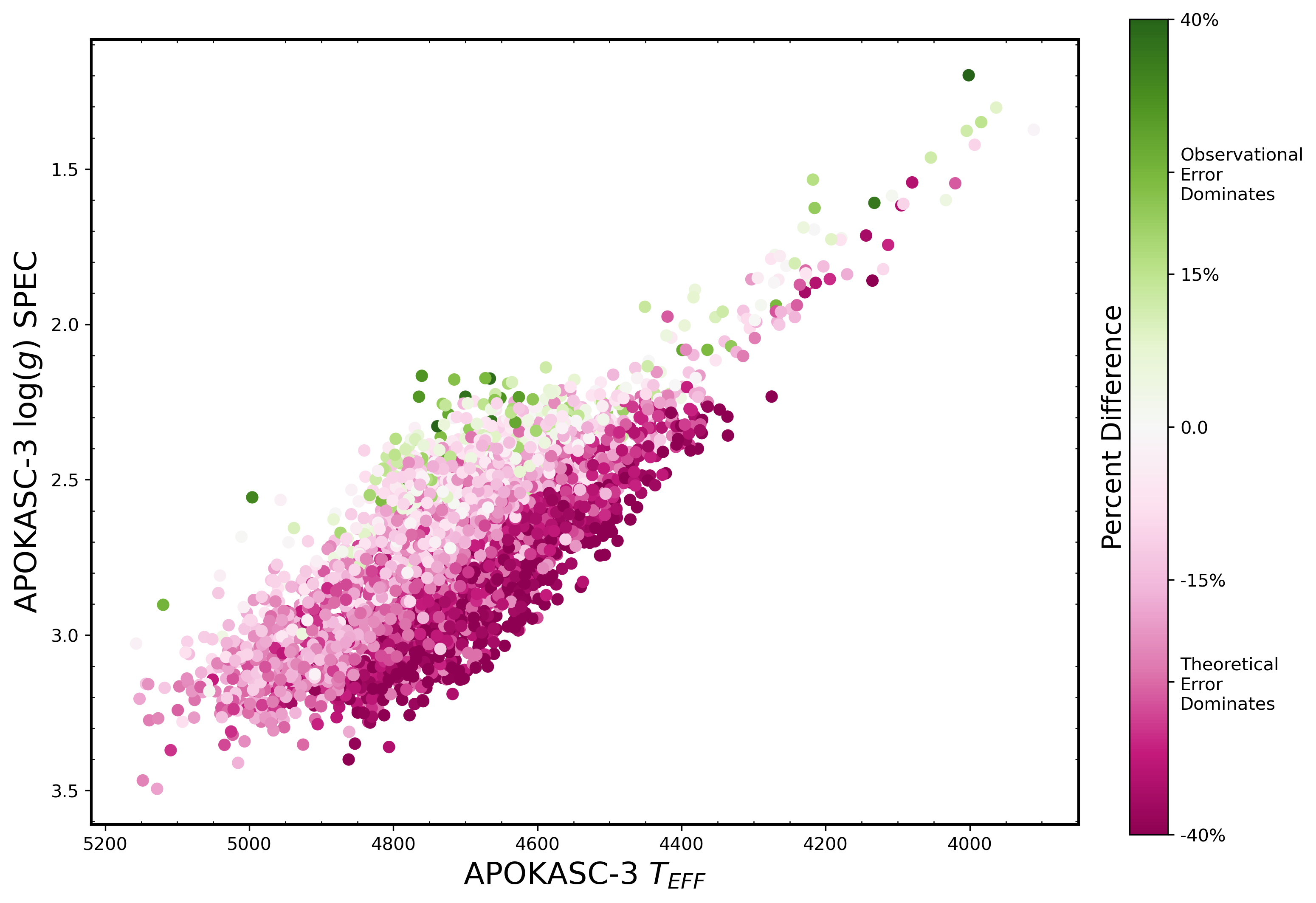}
    \caption{Comparison of observational error to theoretical error resulting from model choice for our spectroscopic interpolation of the APOKASC-3 sample. Our results show that, for the majority of giants, the theoretical error due to model choice exceeds the observational error, with the impact being more pronounced as we decrease in mass.}
    \label{specobserr}
\end{figure}

To understand how the theoretical uncertainty from model choice compares to the age uncertainty from the measurement uncertainties, we followed the same MC procedure as the previous case in Section \ref{aposeis}. Our results in Figure \ref{specobserr} indicate that, for the majority of giants, the theoretical error due to model choice exceeds the observational error, with the impact being more pronounced as we decrease in mass. This implies for those that use ages based solely on spectroscopic data, significant uncertainty inherent in model choice is being overlooked.


\subsubsection{APOKASC-3: Gaia Parameters}
\begin{figure}[hbt]
    \centering
    \includegraphics[scale=.4]{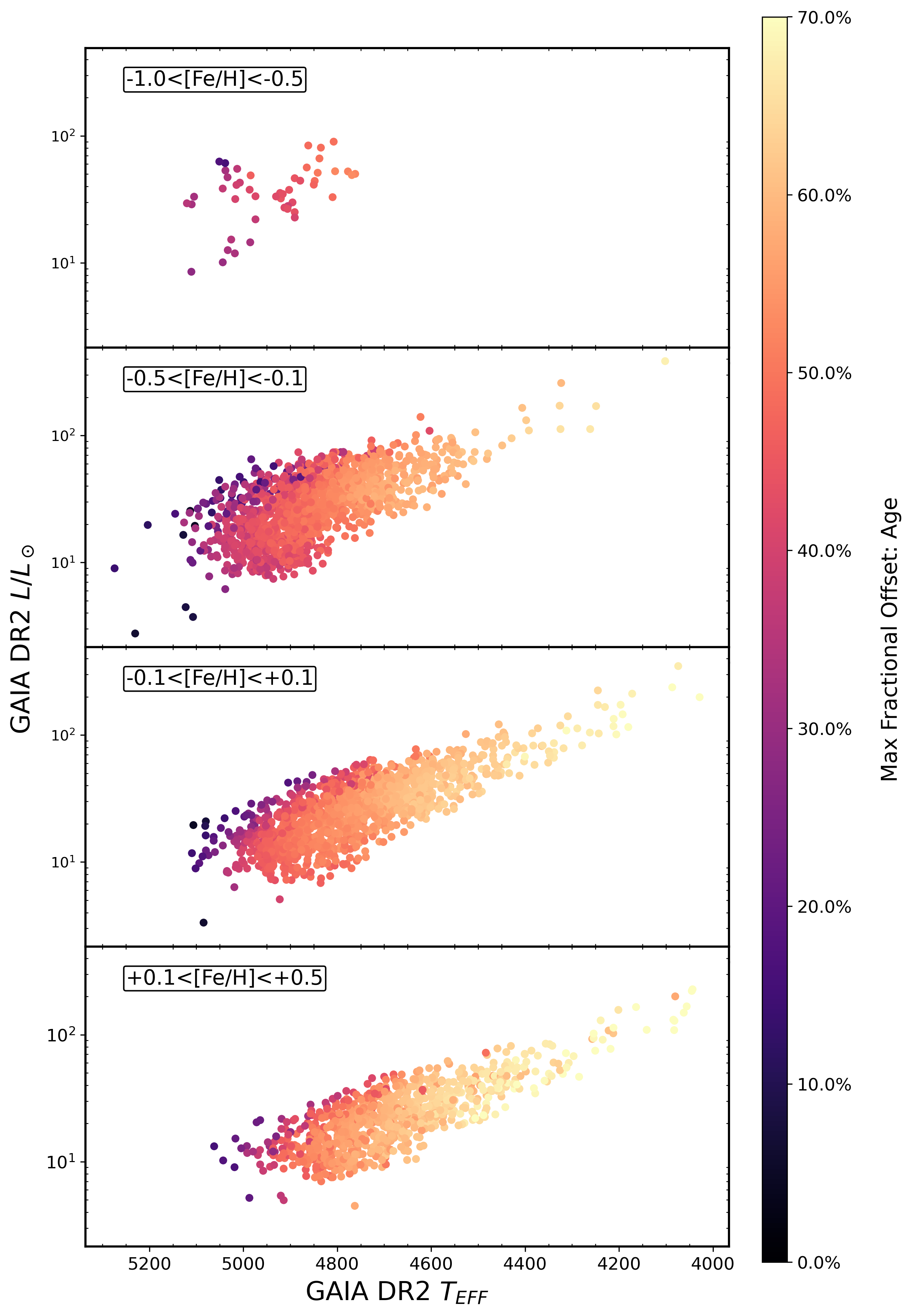}
    \caption{Maximum fractional offset between grids using photometric information from APOKASC-3. We found significant mean offsets to be 42\%, 47\%, 53\%, and 57\% for metallicity regimes $-1.0<[\mathrm{Fe/H}]<-0.5$, $-0.5<[\mathrm{Fe/H}]<-0.$, $-0.1<[\mathrm{Fe/H}]<+0.1$, and $+0.1<[\mathrm{Fe/H}]<+0.5$, respectively. Our results indicate that model grid age inference offsets increase as we ascend the giant branch.}
    \label{gaiagridoffset}
\end{figure}

As a final analysis, we repeated our process using Gaia parameters provided in the APOKASC-3 catalog. We interpolated the grid age estimates using the Gaia DR2 luminosities, effective temperature, and APOKASC-3 spectroscopic metallicities, which have an error on par with Gaia XGBoost metallicities \citep{2023Andrae}. The maximum fractional offset between grids is shown in Figure \ref{gaiagridoffset} and it follows a similar pattern as seen in the spectroscopic comparison, with offsets increasing as we move up the giant branch. We found the average offset in age to be 42\%, 47\%, 53\%, and 57\% for metallicity regimes $-1.0<[\mathrm{Fe/H}]<-0.5$, $-0.5<[\mathrm{Fe/H}]<-0.1$, $-0.1<[\mathrm{Fe/H}]<+0.1$, and $+0.1<[\mathrm{Fe/H}]<+0.5$, respectively. Model agreement improved with Gaia parameters compared to the case with spectroscopic values, though the accuracy remains below that achieved with asteroseismic parameters. We expect that this might mean that the luminosities, which are more directly tied to the nuclear luminosity of the stellar interior, may be a better coordinate for comparison to the models than surface gravity, which depends on the radius and is therefore influenced by the boundary condition and choice in atmosphere. Further investigation into these differences is warranted but lies outside the scope of this project. We show the fractional offset between the APOKASC-3 estimated age and grid age in Figure \ref{fig:ex4} and the mass offset by grid in Figure \ref{fig:ex5}. We calculated the average offset to be 47\% (17\% in mass), 61\% (15\%), 47\% (22\%), and 48\% (25\%) for YREC, GARSTEC, DSEP, and MIST. 

\begin{figure}
    \centering
    \includegraphics[scale=.35]{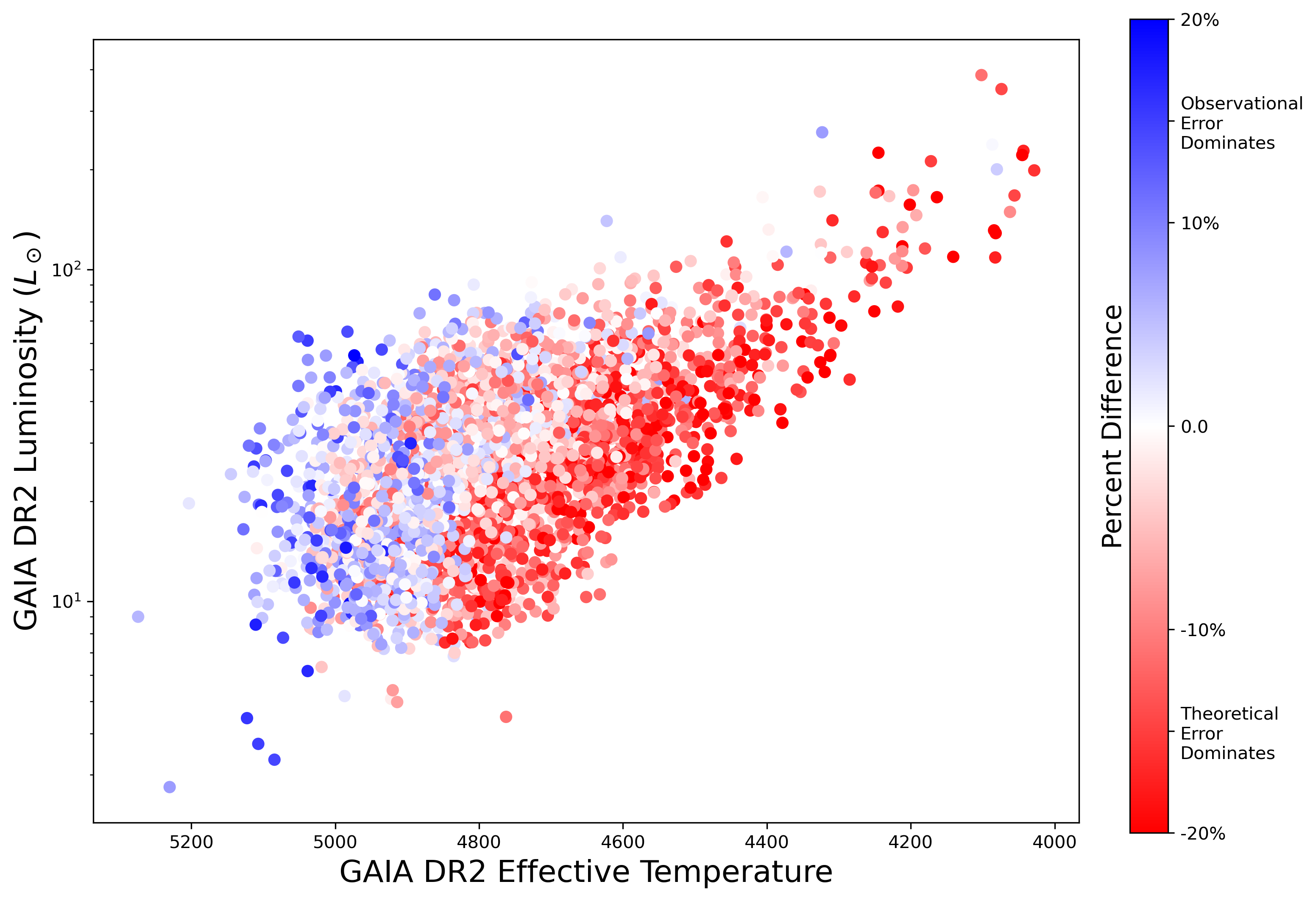}
    \caption{Comparison of observational error to theoretical error resulting from model choice using photometric information from the APOKASC-3 sample. The observational error is found by perturbing the luminosity, metallicity, and effective temperature by $\pm15\%$, $\pm1\sigma$, and $\pm 102$ K, respectively. Our results indicate that theoretical error resulting from model choice outweighs observational error for giants cooler than $\approx 4900$ K, and observational error dominates for evolved giants hotter than $\approx 4900$ K.}
    \label{gaiaobserr}
\end{figure}

To determine the dominant source of uncertainty, we performed a MC analysis, applying the same process used throughout this study. Because the APOKASC-3 catalog does not list uncertainties for the Gaia DR2 parameters,  we adopted literature values for luminosity ($\pm15\%$) and effective temperature $\pm102$ K \citep{2018GaiaDR2}. Figure \ref{gaiaobserr} shows that observational uncertainty outweighs theoretical uncertainty for giants hotter than roughly 4900 K, while theoretical uncertainty becomes dominant for giants cooler than this threshold. Our findings suggest that studies using subgiants for age determination are likely to produce dependable estimates, given the reduced impact of these uncertainties \citep{2022XiangRix}. However, researchers should carefully account for theoretical uncertainties arising from the choice of model grids when deriving ages for cooler giants.

\section{Discussion}\label{discussion}
\subsection{Physics in Question from Previous Studies}\label{modelphysics}
We have shown that different models agree most when mass and metallicity are known, as mass directly affects the nuclear fusion rate and effectively constrains the main sequence (MS) lifetime. The age of a giant is primarily determined by its MS lifetime, and for a star of known mass and composition, the MS lifetime is uncertain by $\sim$10\% percent due to factors such as solar composition scales, convective core overshoot, and the uncertainty on the $^{14}$N(p,$\gamma$) $^{15}$O reaction rate (\citealt{2004Imbriani}; \citealt{2005Weiss}; \citealt{2010Pietrinferni}; \citealt{2018Hidalgo}; \citealt{2022Magg}; \citealt{2022Moedas}). Since the timescale for the giant branch is a small fraction of the MS lifetime, models that agree on mass yield similar age estimates. However, this agreement breaks down significantly when mass is not a known variable. Our demonstrations reveal that disagreements between models increase significantly when we instead infer age from luminosity or surface gravity, and effective temperature. As stated previously, the effective temperature of the RGB is sensitive to changes in the low-temperature opacities, adopted
heavy element distribution, choice of surface boundary conditions, but also in the treatment of convection (\citealt{2002Salaris}; \citealt{2008Vandenberg}; \citealt{2017ApJ...840...17T}; \citealt{2018Salaris}; \citealt{Choi18}). Convection is traditionally parameterized in one-dimensional models using the mixing length theory framework (\citealt{1958BohmVitense}; \citealt{MLT}), simplifying the physics by introducing a mixing length parameter ($\alpha_{MLT}$). Temperature variations along the giant branch are highly sensitive to this semiarbitrary parameter, making such assumptions a source of considerable shifts in the inferred ages of giants.

This sensitivity to convection is also evident in the luminosity of the RGB bump (RGBb), which occurs when the hydrogen-burning shell encounters a mean molecular weight discontinuity left by the deepest extent of the convective envelope (\citealt{2015JCD}; \citealt{2022JTThermoMixing}). As shown in Figure \ref{figure:1}, the luminosity and position of the RGBb vary between models, reflecting differences in convective envelope overshooting treatments and internal mixing processes. While variations in the RGBb complicate interpolation between models, they don’t directly affect age estimates. The luminosity of the RGBb remains a topic of ongoing discussion, with recent studies exploring the impacts of mixing and the depth of the convective layer on its luminosity (\citealt{2018Khan}; \citealt{2020VSA}; \citealt{2022Lindsay}; \citealt{2023Miller}). 

\subsection{Tools}
Our goal is not only to bring light to the caveats of stellar modeling on the RGB, but also to provide a method that aids in estimating the age of red giants by utilizing multiple stellar models that account for different internal processes, resulting in more realistic age estimations and uncertainties. To support this, we provide an additional Jupyter Notebook{\footnote{\url{https://github.com/zclaytor/kiauhoku/blob/main/notebooks/model_offsets_RGB.ipynb}}} interface that allows the reader to explore {\fontfamily{qcr}\selectfont kiauhoku} and interpret the inferred information from the models. The notebook includes examples that demonstrate how to interpolate ages based on the different parameter combinations we analyzed, enabling users to reproduce our approach. In addition, we include plots that allow readers to confirm that the characteristics of their star match the most suitable model evolutionary track, helping to ensure the reliability of the model-derived ages. 

Because the EEP spacing in the YREC, GARSTEC, and DSEP grids used in this study differs from that in our main version of {\fontfamily{qcr}\selectfont kiauhoku}, we have made the additional grids available on Zenodo{\footnote{\url{https://zenodo.org/records/14908017}}} for readers who wish to reproduce our results.

\subsection{Improvement and Impacts}
Our study highlights that specific aspects of stellar internal processes require further attention to better characterize evolutionary paths. While constraining stellar physics presents significant challenges, a more detailed analysis is necessary to determine which physical factors are most impactful and in which contexts. We refrain from asserting the superiority of one model over another, as each makes justifiable choices in stellar physics, and most models have a particular regime for which they were designed and at which they are best. Instead, we take a more global view and argue that researchers should exercise caution when using evolutionary tracks to estimate the masses of giants, as these estimates can vary significantly depending on the model used. Additionally, studies involving cool, evolved giants should be approached with care, as offsets increase in these regions. We argue that in cases where either absolute or relative ages matter, age inference uncertainties should, at a minimum, account for the variation across multiple grids to better reflect the ongoing studies of stellar processes. 

\section{Conclusion}
Given the essential role of precise stellar ages in galactic archaeology, where they are crucial for constructing accurate timelines of the evolution of our Universe, the goal of this study was to refine age estimates and uncertainties for red giants through the use of evolutionary models. With advancements in spectroscopic and photometric surveys, we have reached a point where theoretical uncertainties from model choice now exceed observational errors for red giants. Although this currently presents a challenge for precise observational inference, it also provides an opportunity to rigorously test and refine our stellar models. We summarize our concluding results below. 
\begin{itemize}
    \item For readers using asteroseismic parameters or other methods that provide a precise mass, our results show that the age estimates across the grids are in agreement, with a mean offset of 10\%.
    \item When mass is not known and must instead be inferred solely from spectroscopic or Gaia data, using parameters such as luminosity, surface gravity, metallicity, and effective temperature, the theoretical uncertainty in age due to model choice becomes significantly larger, reaching a mean offset of 90\%. 
    \item We found that when the asteroseismic parameters are used for grid age interpolation, the shift in age estimations given the observational uncertainties outweighs the theoretical uncertainties from model choice. In contrast, when spectroscopic parameters are used, the theoretical uncertainty resulting from model choice exceeds the shift in age due to measurement uncertainty for most giants. When using Gaia parameters, theoretical uncertainty is greater than the observational uncertainty in giants cooler than $\approx 4900$K, while the opposite is true for evolved giants hotter than $\approx4900$K. 
\end{itemize}

The expansion of high-precision stellar surveys, such as Gaia \citep{2023Gaia}, SDSS \citep{SDSSV}, and the upcoming Roman (\citealt{2015Roman}; \citealt{2019Roman}) and Rubin LSST \citep{2019LSST} missions, is providing an abundance of data, enabling us to ask increasingly detailed questions about the history and evolution of our galaxy \citep{2024Deason}. Such advancements also facilitate connections to other galaxies \citep{2023Wang} and models of galactic  structure and dynamics \citep{2023Binney}. However, many of these inferences rely on either precise or accurate ages \citep{2024Gozaliasl}. Therefore, we emphasize the importance of accounting for all sources of uncertainties, including those arising from both observation and theory. 

\section*{Acknowledgements}
L.M. and J.T. acknowledge support from NASA grant 80NSSC22K0812. Z.R.C. acknowledges support from NASA grant 80NSSC24K0081. The authors thank the anonymous referee for their insightful feedback and careful review of this work.

\clearpage
\bibliography{referencesJT, referencesmine,msJT,ms2JT}{}
\bibliographystyle{aasjournal}

\clearpage 

\appendix

\renewcommand{\thefigure}{A.\arabic{figure}} 
\setcounter{figure}{0} 
\section{APOKASC-3} 
\label{sec:appendix}

\subsection{Age inference Using Asteroseismic Data}
\begin{figure}[hbt]
    \centering
    \includegraphics[scale=.65]{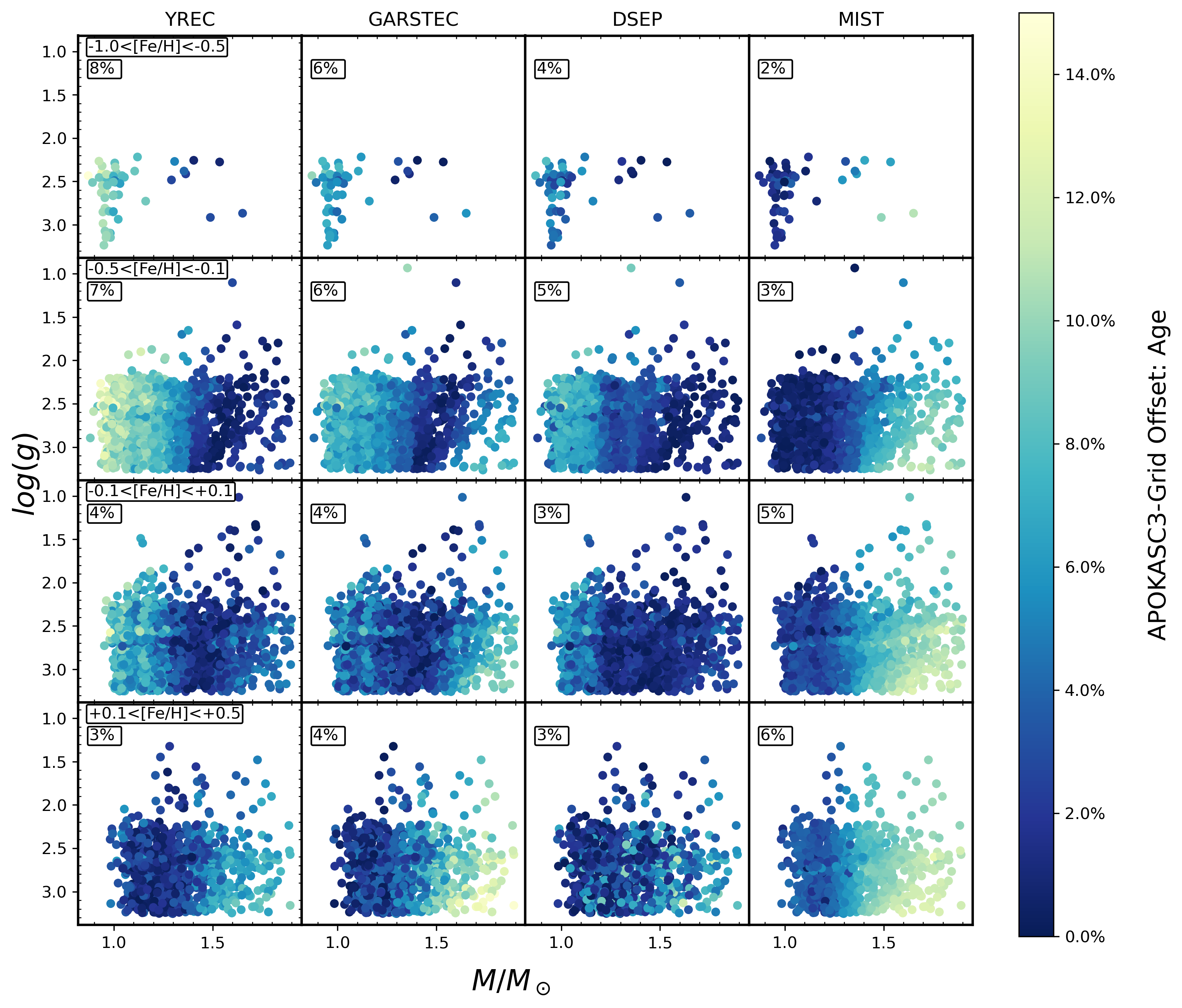}
    \caption{Mass vs surface gravity for red giants in the APOKASC-3 sample. We show the offset between APOKASC-3 ages and the grid ages using asteroseismic parameters, organized in metallicity bins. Starting from the top panels, the metallicity ranges are as follows: $-1.0<[\mathrm{Fe/H}]<-0.5$, $-0.5<[\mathrm{Fe/H}]<-0.1$, $-0.1<[\mathrm{Fe/H}]<+0.1$, and $+0.1<[\mathrm{Fe/H}]<+0.5$. The age estimates are interpolated using the APOKASC-3 asteroseismic mass, and spectroscopic metallicity and surface gravity. We find mean average offsets from the ages inferred with the APOKASC-3 models of 5\%, 5\%, 4\%, and 4\% for YREC, GARSTEC, DSEP, and MIST, respectively. The mean age offset for each regime is displayed within each panel.}
    \phantomsection
    \label{fig:ex1}
\end{figure}
\clearpage
\subsection{Age inference Using Spectroscopic Data}
\begin{figure}[hbt]
    \centering
    \includegraphics[scale=.65]{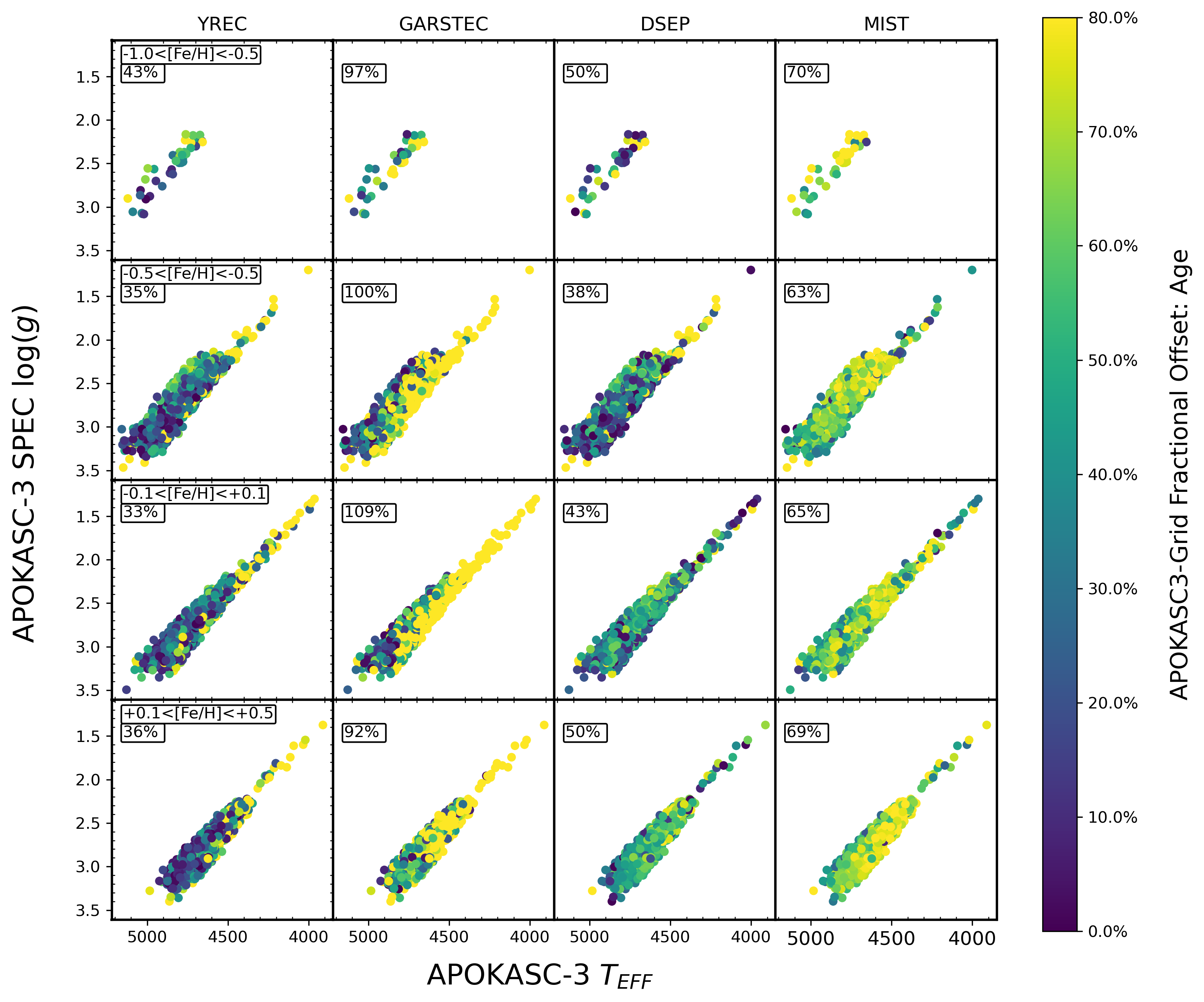}
    \caption{APOKASC-3 effective temperature vs surface gravity for red giants in the APOKASC-3 sample, color coded by the offset between the APOKASC-3 age inferred using asteroseismic parameters and the age inferred using spectroscopic measurements for each model grid, organized into metallicity bins. We used the spectroscopic data, surface gravity, metallicity and effective temperature from APOKASC-3. Starting from the top panels, for metallicity ranges  $-1.0<[\mathrm{Fe/H}]<-0.5$, $-0.5<[\mathrm{Fe/H}]<-0.1$, $-0.1<[\mathrm{Fe/H}]<+0.1$, and $+0.1<[\mathrm{Fe/H}]<+0.5$, we find mean offsets in age of 37\%, 100\%, 45\%, and 67\% for YREC, GARSTEC, DSEP, and MIST, respectively. The mean offset for each regime is displayed within each panel.}
    \phantomsection
    \label{fig:ex2}
\end{figure}
\clearpage
\begin{figure}[hbt]
    \centering
    \includegraphics[scale=.65]{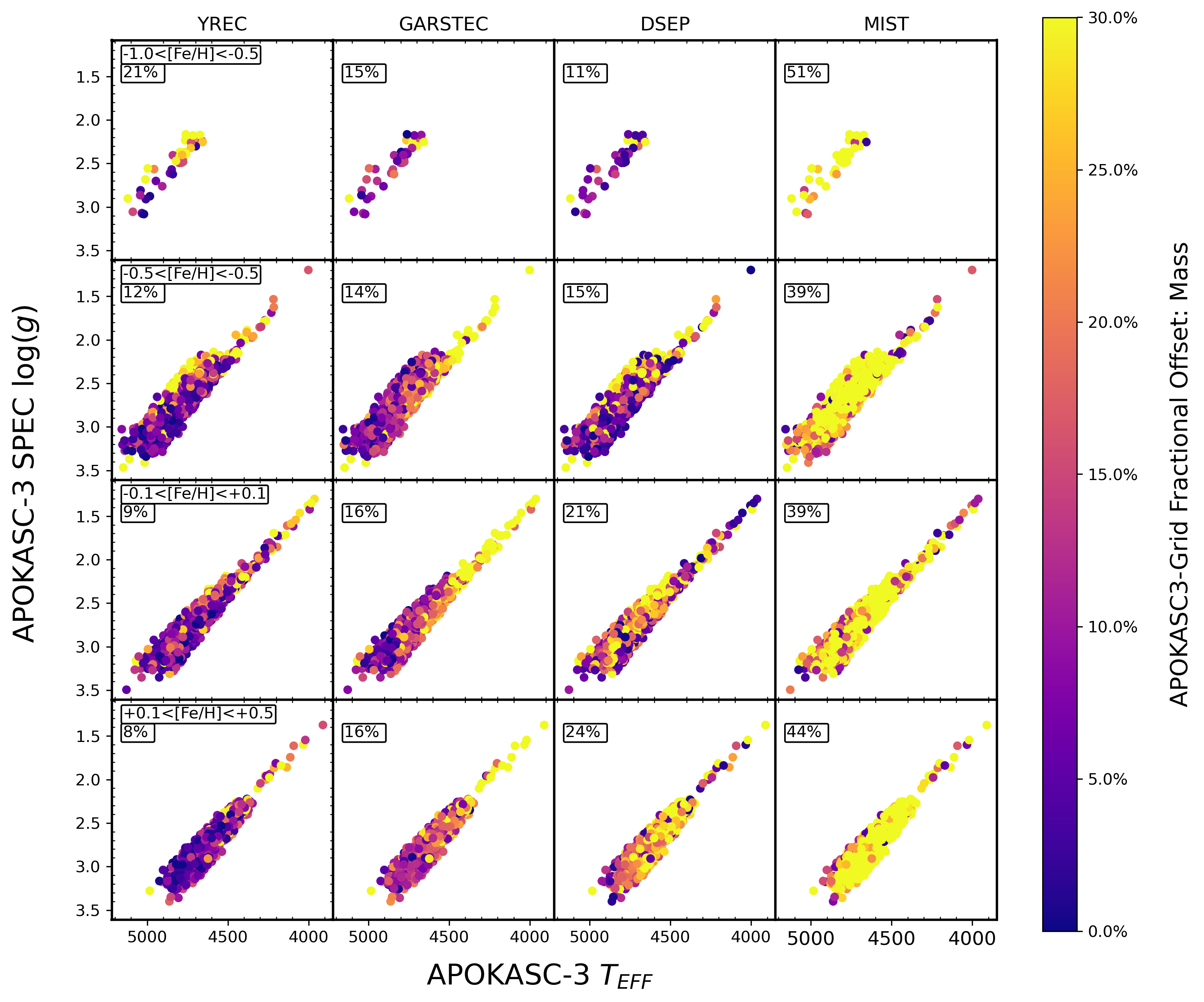}
    \caption{APOKASC-3 effective temperature vs surface gravity for red giants in the APOKASC-3 sample, color coded by the offset between APOKASC-3 asteroseismic mass and the grid-estimated mass found using spectroscopic data, organized into metallicity bins. We find mean offsets in mass to be 12\%, 15\%, 18\%, and 43\% for YREC, GARSTEC, DSEP, and MIST, respectively. The mean offset for each regime is displayed within each panel.}
    \phantomsection
    \label{fig:ex3}
\end{figure}

\clearpage

\subsection{Age inference Using Gaia Photometric Data}
\begin{figure}[hbt]
    \centering
    \includegraphics[scale=.65]{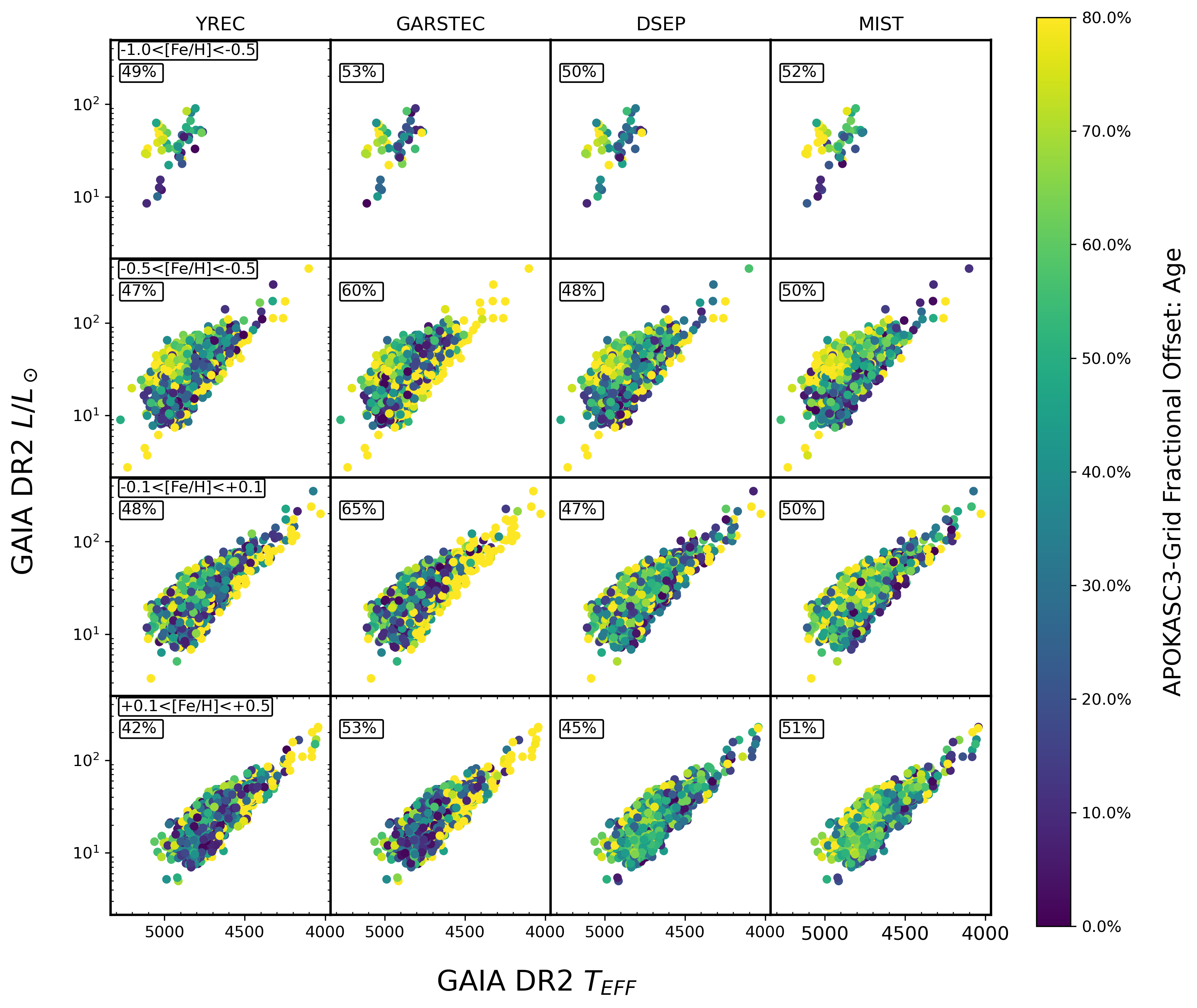}
    \caption{GAIA DR2 effective temperature vs GAIA DR2 luminosity for red giants in the APOKASC-3 sample, color coded by the offset between the APOKASC-3 age inferred using asteroseismic parameters and the age inferred using photometric measurements for each model grid, organized into metallicity bins. For each grid-age estimate, we used Gaia luminosity and effective temperature, and spectroscopic metallicity from APOKASC-3. Starting from the top panels, for metallicity ranges $-1.0<[\mathrm{Fe/H}]<-0.5$, $-0.5<[\mathrm{Fe/H}]<-0.1$, $-0.1<[\mathrm{Fe/H}]<+0.1$, and $+0.1<[\mathrm{Fe/H}]<+0.5$, we find mean offsets of 47\%, 61\%, 47\%, and 48\% for YREC, GARSTEC, DSEP, and MIST, respectively. The mean offset for each regime is displayed within each panel.}
    \phantomsection
    \label{fig:ex4}
\end{figure}

\begin{figure}[hbt]
    \centering
    \includegraphics[scale=.65]{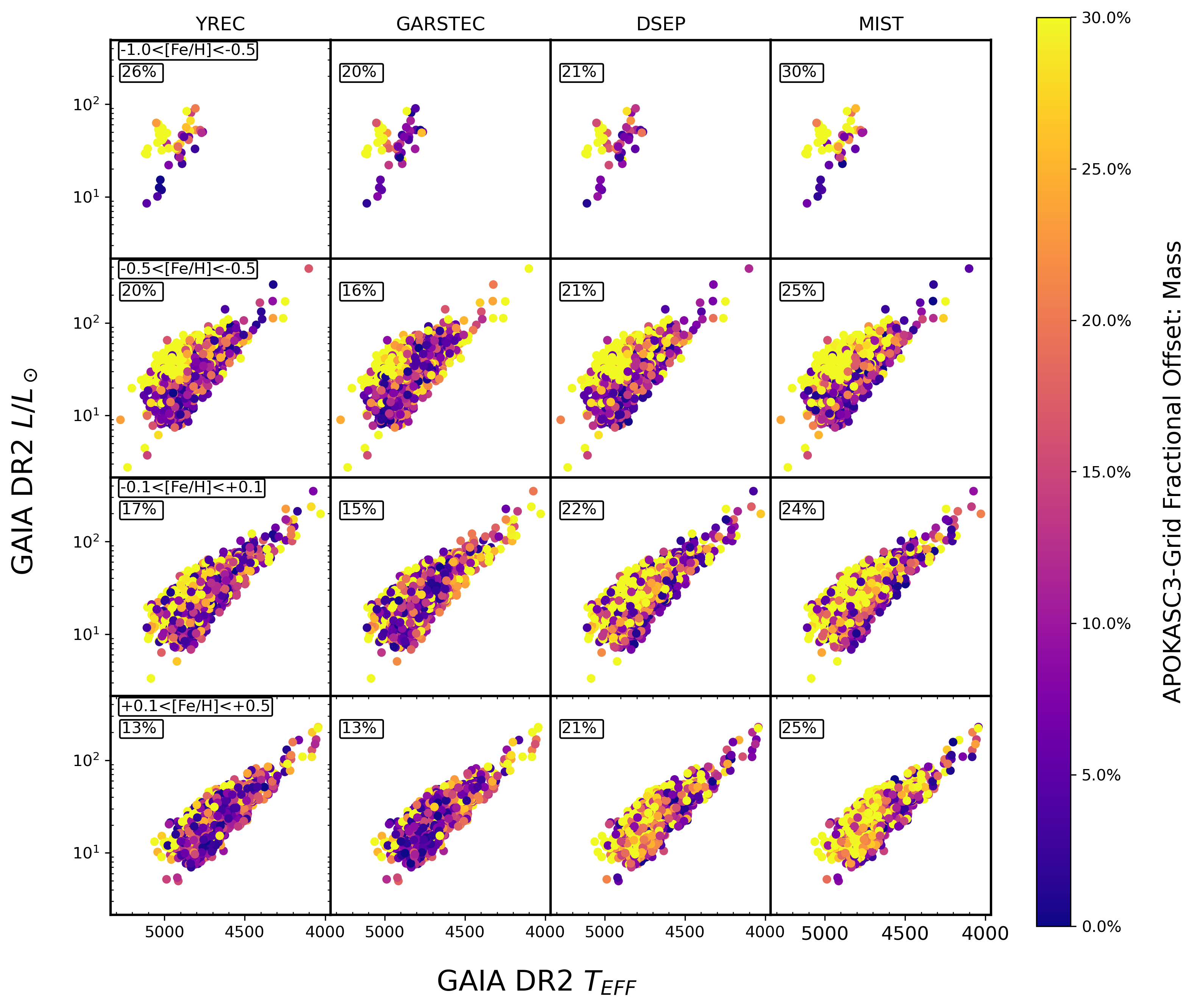}
    \caption{GAIA DR2 effective temperature vs GAIA DR2 luminosity for red giants in the APOKASC-3 sample, color coded by the offset between APOKASC-3 asteroseismic mass and the grid-estimated mass found using Gaia data, organized into metallicity bins. We find mean offsets in mass to be 17\%, 15\%, 22\%, and 25\% for YREC, GARSTEC, DSEP, and MIST, respectively. The mean offset for each regime is displayed within each panel.}
    \phantomsection
    \label{fig:ex5}
\end{figure}
\clearpage
\subsection{Monte Carlo Simulation Results}

\begin{figure}[hbt]
    \centering
    \includegraphics[scale=.45]{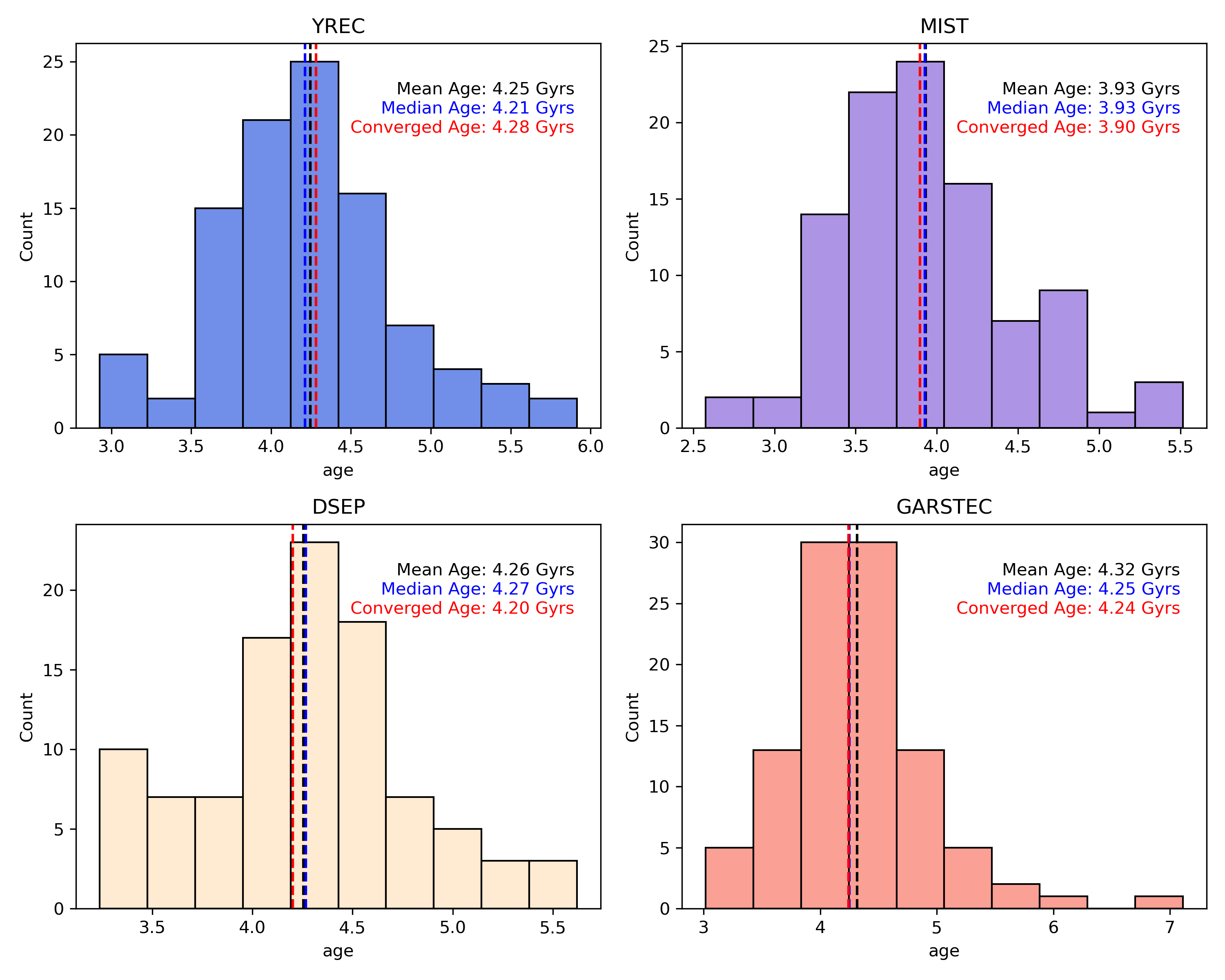}
    \caption{Histogram of stellar ages from the MC of 100 iterations for star KIC: 10023322 using the APOKASC-3 mass, spectroscopic metallicity, and asteroseismic effective temperature. The distribution represents the uncertainty due to observational error. In the relevant sections of the paper, we report the age spread encompassing 50\% of the probability distribution as the observational uncertainty.}
    \phantomsection
    \label{fig:mcexmass}
\end{figure}

\begin{figure}[hbt]
    \centering
    \includegraphics[scale=.45]{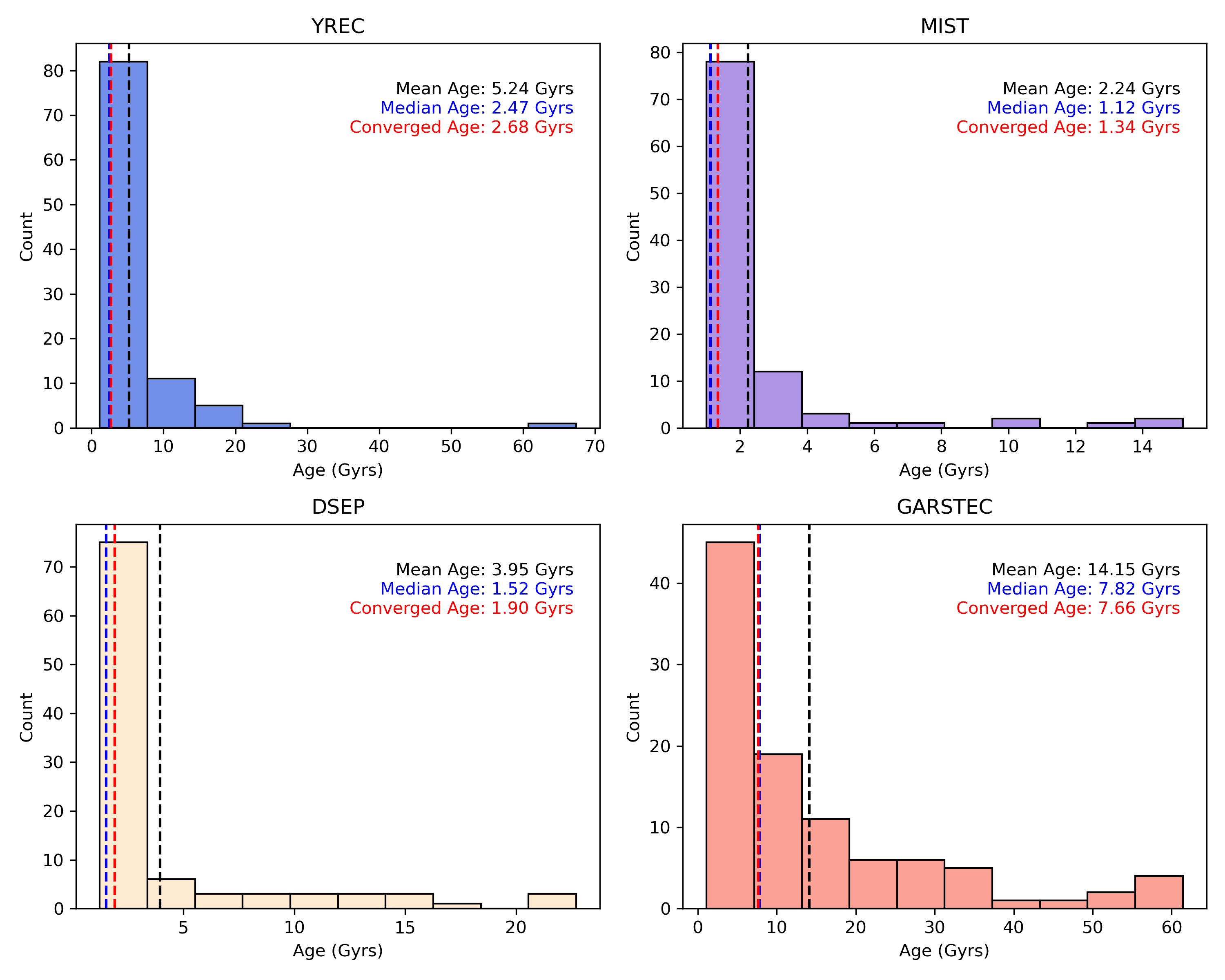}
    \caption{Histogram of stellar ages from the MC of 100 iterations for star KIC: 10023322 using the APOKASC-3 spectroscopic surface gravity, metallicity, and effective temperature. The distribution represents the uncertainty due to observational error.}
    \phantomsection
    \label{fig:mcexspec}
\end{figure}

\begin{figure}[hbt]
    \centering
    \includegraphics[scale=.45]{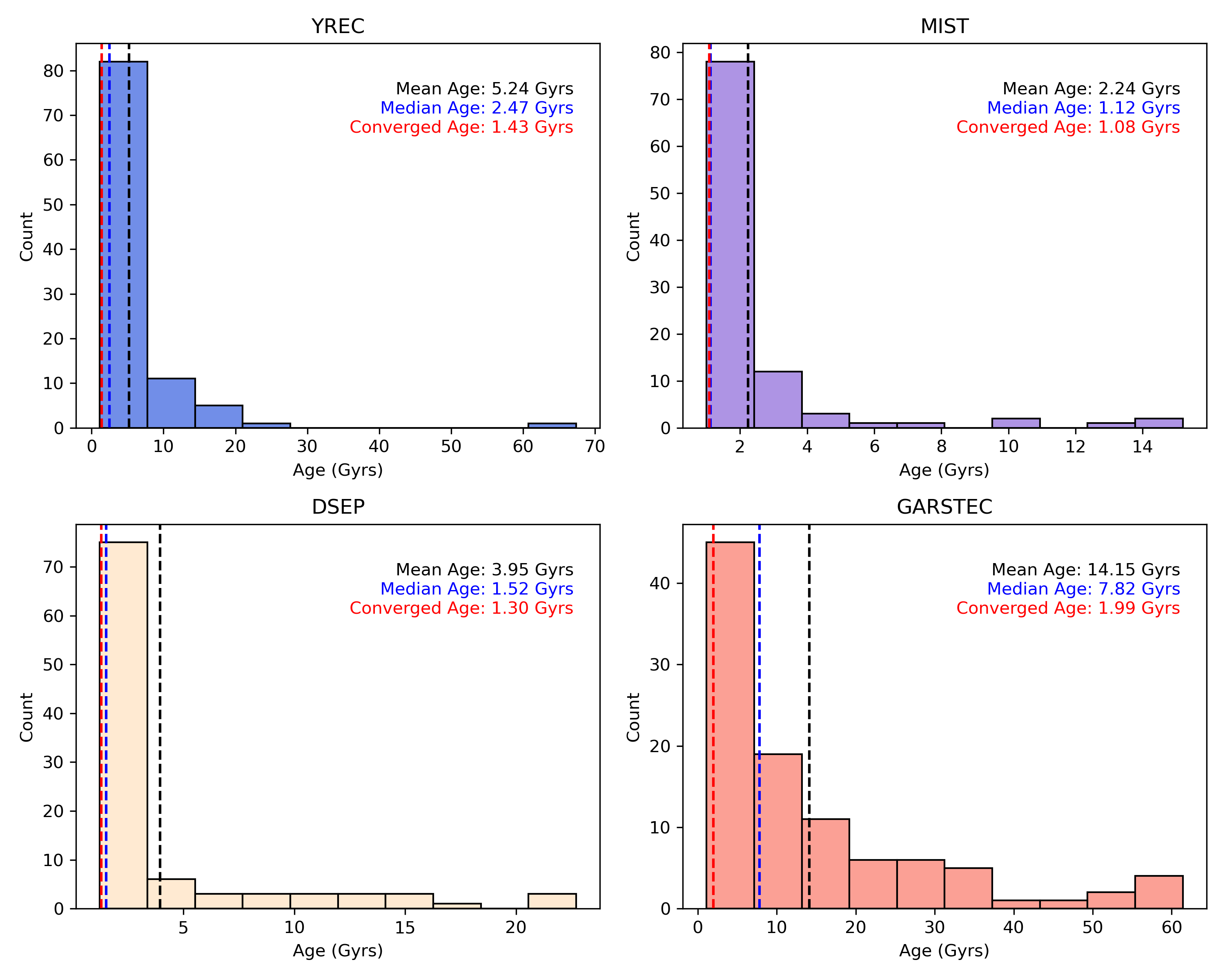}
    \caption{Histogram of stellar ages from the MC of 100 iterations for star KIC: 10023322 using the Gaia DR2 luminosity, effective temperature, and APOKASC-3 spectroscopic metallicity. The distribution represents the uncertainty due to observational error.}
    \phantomsection
    \label{fig:mcexgaia}
\end{figure}

\end{document}